%% file: paper.tex
\documentclass[lettersize,journal]{IEEEtran}
\usepackage{tex/packages}
\def\BibTeX{{\rm B\kern-.05em{\sc i\kern-.025em b}\kern-.08em
    T\kern-.1667em\lower.7ex\hbox{E}\kern-.125emX}}

\addto\captionsenglish{}

\begin{document}
\input{acronyms}

\title{A Control-Bounded Quadrature Leapfrog ADC}
\author{Hampus Malmberg~\IEEEmembership{Member,~IEEE}, Fredrik Feyling,~\IEEEmembership{Graduate~Student~Member,~IEEE},\\and Jos\'{e} M. de la Rosa,~\IEEEmembership{Fellow,~IEEE}
\thanks{Manuscript submitted October 27, 2023;

Hampus Malmberg is with the Department of Information Technology \& Electrical Engineering, ETH Zürich, 8092 Zürich, Switzerland (email: malmberg@isi.ee.ethz.ch)

Fredrik Feyling is with Department of Electronic Systems, Norwegian University of Science and Technology, 7491 Trondheim, Norway (email: fredrik.e.feyling@ntnu.no)

Jos\'{e} M. de la Rosa is with the Institute of Microelectronics of Seville, IMSE-CNM (CSIC/University of Seville), 41092 Sevilla, Spain (e-mail: jrosa@imse-cnm.csic.es)

}}
\IEEEpubid{This work has been submitted to the IEEE for possible publication. Copyright may be transferred without notice, after which this version may no longer be accessible.}
\maketitle

\begin{abstract}
In this paper, the design flexibility of the control-bounded analog-to-digital converter principle is
demonstrated. A band-pass analog-to-digital converter is considered as an application and case study. We show how a low-pass control-bounded analog-to-digital converter can be translated into a band-pass version where the guaranteed stability, converter bandwidth, and signal-to-noise ratio are preserved while the center frequency for conversion can be positioned freely. The proposed converter is validated with behavioral simulations on several filter orders, center frequencies, and oversampling ratios. Additionally, we consider an op-amp circuit realization where the effects of first-order op-amp non-idealities are shown. 
Finally, robustness against component variations is demonstrated by Monte Carlo simulations.
\end{abstract}


\section{Introduction}

\IEEEPARstart{B}{and}-Pass Sigma-Delta Modulators (BP-$\Sigma\Delta$Ms) and \acp{QSDM} allow digitizing non-base-band signals; essentially 
expediting the role and position of \ac{A/D} conversion 
in the receiving structure of wireless receivers.
A mainly digital wireless receive path is beneficial as digital signal processing offers better technology scaling and programmability towards a
\ac{SDR} platform \cite{saye20, ghae21, jie21}.
Digitizing \ac{RF} signals requires sampling frequencies in the GHz range. Hence, state-of-the-art \ac{BPSDM}s are primarily implemented using \ac{CT} circuits as they offer inherent anti-aliasing filtering and are potentially faster and more power efficient than their \ac{DT} counterparts. However, in the majority of cases, \ac{RF} \acp{BPSDM} have a fixed ratio between the center or \textit{notch} frequency, $f_n$, and the sampling frequency $f_s$ (typically $f_n=f_s/4$).
A fixed $f_s/f_n$ ratio results in two main limitations: firstly, for wireless standards operating around 2.5-5GHz, prohibitive values of $f_s$, in the order of tens of GHz, are typically required. Secondly, a widely programmable \ac{PLL} is required for tuning $f_n$ while keeping the $f_s/f_n$ ratio fixed.
These limitations have prompted the interest in reconfigurable \ac{BPSDM}s with tunable notch
frequency \cite{moli14,Shib12}. However, reported solutions are limited in practice by the increased (analog) circuit complexity and risk of the potential instability of the loop filter -- compromised by the tuning range of $f_n$\cite{moli14}. \acp{QSDM} has the additional challenge of quadrature errors caused by mismatch between the I and Q channels of the modulator \cite{ST:05}.

This paper extends upon the work in \cite{MFR:23} and presents an alternative approach to the problem of digitizing \ac{RF} signals using the so-called \ac{CBADC} concept \cite{M:20,MWL:21}. A \ac{QCBADC} is proposed, that offers a highly modular architecture with a tunable $f_n$ and an analytical stability guarantee. In particular, the \ac{QCBADC} follows from extending two low-pass \acp{CBADC} into a single oscillating structure. Conveniently, the \ac{QCBADC}'s \ac{SNR} and \ac{BW} specification follows from its two low-pass \acp{CBADC} building blocks.
Like \acp{QSDM}, the \acp{QCBADC} is a quadrature \ac{ADC}, resulting in the same number of integrating stages per signal as its low-pass building block. 

The manuscript is organized as follows: \Sec{sec:fundamentals} gives a brief introduction to the \ac{CBADC} framework and its relation to \acp{CTSDM}.
\Sec{sec:leapfrog} and \Sec{sec:general_digital_estimator} generalizes beyond the \ac{CTSDM} example and introduces the low-pass leapfrog \ac{CBADC} structure.
\Sec{sec:quadrature} demonstrates how a low-pass \ac{CBADC} can be transformed into a \ac{QCBADC}. To exemplify the \ac{QCBADC}, \Sec{sec:leap_frog_quadrature} shows the quadrature version of the leapfrog \ac{CBADC} from \Sec{sec:leapfrog} and validates the presented approach by electrical simulations including circuit non-idealities. \Sec{sec:conlcusions} concludes our findings. 
Finally, to enhance readability, longer derivations from the main text are presented in the Appendix.

\IEEEpubidadjcol
\section{\ac{CBADC} Fundamentals}\label{sec:fundamentals}
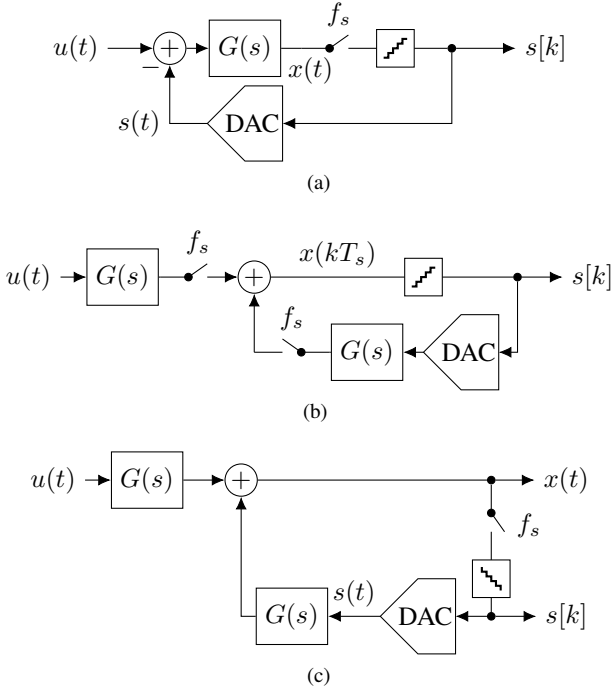
\begin{figure}
    \centering
    \subfloat[\label{fig:ct-modulator}]{%
       \input{figures/sigma_delta_ct}
    } \\
     \subfloat[
    \label{fig:sdm_est_2}]{%
            \input{figures/sigma_delta_estimation_model_2}
    } \\
    \subfloat[\label{fig:sdm_est_1}]{%
            \input{figures/sigma_delta_permutation_1}
    }
    \caption{\label{fig:ctsdm}%
    \protect\subref{fig:ct-modulator} The \ac{CTSDM}. 
    \protect\subref{fig:sdm_est_2} The \ac{CTSDM} rearranged to emphasize the modulator from a discrete-time perspective, which is central to the \ac{CTSDM} estimation principle in \Sec{sec:ctsdm_estimation}.
    \protect\subref{fig:sdm_est_1} The \ac{CTSDM} rearranged to emphasize the bounded continuous-time observation $x(t)$ which is central to the \ac{CBADC} estimation principle in \Sec{sec:cbadc_est}.
    }
\end{figure}

The \ac{CBADC} concept originates from statistical signal processing concepts \cite{LDHKPK:2007, FL:2007, BL:2008, LBWB:11, BLV:2013, BL:2014, LW:15} and was developed \cite{M:20, MWL:21} independently of \acp{CTSDM}. 
Nevertheless, the two concepts undeniably share several commonalities. In fact, as will be shown in \Sec{sec:cbadc_est}, any \ac{CTSDM} can be viewed as a \ac{CBADC}. 
As \ac{CTSDM} is a well-established and generally understood concept within the \ac{ADC} community, we will demonstrate the \ac{CBADC} concept using the general \ac{CTSDM} in \Fig{fig:ct-modulator}. 

The \ac{CTSDM}
takes an analog continuous-time input signal $u(t)$ and produces a discrete-time digital output $s[k]$. The modulator includes an analog continuous-time loop filter $G(s)$, a sampling stage operating at $f_s$ samples-per-second, a quantizer, and a \ac{DAC}. We define the internal signals
\begin{IEEEeqnarray}{rCl}
    s(t) & \eqdef & \sum_{k\in\Z} \mathrm{DAC}(t - k T_s) s[k], \label{eq:ct_control_signal}
\end{IEEEeqnarray}
for $T_s \eqdef 1 / f_s$,
the analog continuous-time loop filter's output signal as $x(t)$,
and the quantization error as
\begin{IEEEeqnarray}{rCl}
    q[k] & \eqdef & s[k] - x(k T_s). \label{eq:quantization_signal}
\end{IEEEeqnarray}

\subsubsection{Conventional \ac{CTSDM} Estimation}\label{sec:ctsdm_estimation}
\begin{figure*}
    \centering
    \subfloat[\label{fig:sdm_est}]{%
            \input{figures/sigma_delta_discrete_time}
    } \\
    \subfloat[\label{fig:est_cbadc}]{
            \input{figures/cbadc_estimation}
    }
    \caption{\label{fig:estimation_model} 
    \protect\subref{fig:sdm_est} The \ac{CTSDM} estimation model of \Fig{fig:ctsdm}, from \Sec{sec:ctsdm_estimation}, when conditioning on $U(i\omega)$ being band-limited through $G(i\omega)$.
    \protect\subref{fig:est_cbadc}  The \ac{CBADC}'s estimation model of \Fig{fig:ctsdm}, from \Sec{sec:cbadc_est}, when $X(i\omega)$ is assumed to be an independent variable.
    }
\end{figure*}
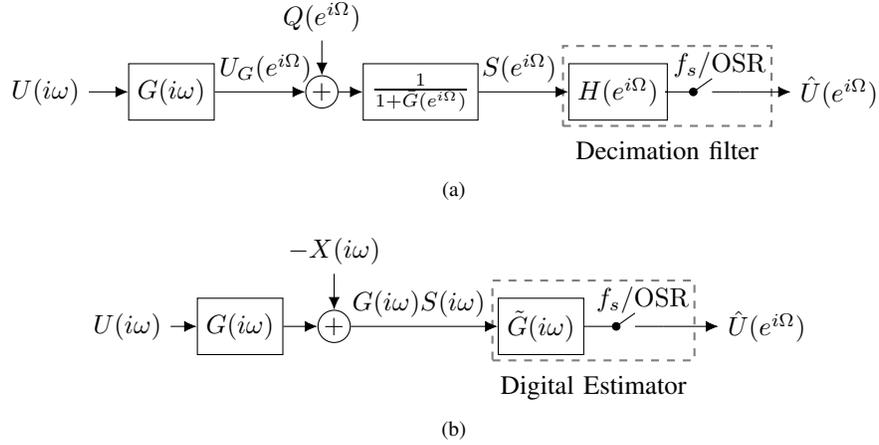

The conventional estimation approach, cf. \cite{R:2011, OG:2006}, 
is to emphasize the discrete-time aspects of the \ac{CTSDM} operation by rearranging \Fig{fig:ct-modulator} as in \Fig{fig:sdm_est_2}. As for discrete-time sigma-delta modulators (DT-$\Delta\Sigma$Ms),
the quantization step is modeled as an additive quantization error \Eq{eq:quantization_signal}.
Finally, by assuming \Eq{eq:quantization_signal} to be an independent statistical noise process and the input signal $u(t)$ to be band-limited through $G(s)$,
follows the estimation model in \Fig{fig:sdm_est} for
\begin{IEEEeqnarray}{rCl}%
    U_G(e^{i \Omega}) 
    & = & G(i\omega)U(i\omega) \\
    \bar{G}\left(e^{i \Omega}\right) & \eqdef & \sum_{k \in \Z} G\left(i\left(\omega + 2 \pi k f_s \right) \right)\mathrm{DAC}\left(i\left( \omega + 2 \pi k f_s\right)\right) \nonumber \\
\end{IEEEeqnarray}
where $|\omega T_s| = |\Omega| < \pi$.
As the modulator output, $S(e^{i\Omega})$, contains out-of-band noise, it is customary to, at least conceptually, formulate the final estimate $\hat{U}(e^{i\Omega})$ as the down-sampled output of a tentative decimation filter $H(e^{i\Omega})$.

\subsubsection{\ac{CBADC} Estimation} \label{sec:cbadc_est}
The \ac{CBADC} estimation approach
emphasizes the continuous-time aspects of the \ac{CTSDM} operation by rearranging \Fig{fig:ct-modulator} as in \Fig{fig:sdm_est_1}. In this view, $s[k]$ is not considered a sampled and quantized observation of a filtered input signal, but instead 
\begin{IEEEeqnarray}{rCl}
        x(t) & = & (g \ast u)(t) - (g \ast s)(t) \label{eq:ct_observation}
\end{IEEEeqnarray}
is the principal observation 
where $g(t)$ is the impulse response of $G(s)$ and $\ast$ refers to continuous-time convolution. In words, \Eq{eq:ct_observation} states that $(g \ast s)(t)$,
which follows from $g(t)$, $\mathrm{DAC}(t)$, and $s[k]$,
is an arbitrarily good continuous-time approximation of $(g \ast u)(t)$ for a relatively large open-loop-gain $G(s)$ in relation to a bounded loop-filter output $x(t)$.

Thinking of bounded continuous-time observations
may appear esoteric. 
Concretely, \Eq{eq:ct_observation} connects, the discrete-time convolution, 
\begin{IEEEeqnarray}{rCl}
    \hat{u}[k] & \eqdef & 
    \sum_{k_1\in \Z} h[k_1] s[k - k_1] \label{eq:ct_estimate}
\end{IEEEeqnarray}
with the sampling operation
\begin{IEEEeqnarray}{rCl}
    \hat{u}[k] & = & (\tilde{g} \ast g \ast u)(k T_s) - (\tilde{g} \ast x)(kT_s) \label{eq:ctsdm_decomp}
\end{IEEEeqnarray}
for
\begin{IEEEeqnarray}{rCl}
    h[k] & \eqdef &  (\tilde{g} \ast g \ast \mathrm{DAC})(k T_s). \label{eq:ct_filter_coefficients}
\end{IEEEeqnarray}

To avoid confusion, we stress that selecting a discrete-time filter $h[k]$ and implementing \Eq{eq:ct_estimate} 
is the practical task of the \ac{CBADC} digital estimator, whereas the sampling in \Eq{eq:ctsdm_decomp} 
and the impulse response $\tilde{g}(t)$ in \Eq{eq:ct_filter_coefficients} follows implicitly from the choice of $h[k]$, $G(s)$, and $\mathrm{DAC}(s)$.
The derivation of \Eq{eq:ctsdm_decomp} from \Eq{eq:ct_estimate} are given in \App{app:decomposition_scalar_case}. 

Note that, as the \ac{CTSDM} oversamples by a \ac{OSR} factor, \Eq{eq:ct_estimate} is typically only evaluated for every \ac{OSR} sample.

The sampling perspective \Eq{eq:ctsdm_decomp} is a helpful design tool in relating the estimation error to the loop-filter $G(s)$ and the remainder signal $x(t)$. In particular, when assuming $x(t)$ to be an independent variable, the corresponding estimation model can be illustrated as in \Fig{fig:est_cbadc} where $X(s)$ and $\tilde{G}(s)$ refer to the Laplace transforms of $x(t)$ and $\tilde{g}(t)$ respectively.
Furthermore, if $h[k]$ is chosen such that $\tilde{G}(s)$ band-limits both terms in \Eq{eq:ctsdm_decomp}, we have
\begin{IEEEeqnarray}{rCl}
    \hat{U}\left(e^{i \Omega}\right) & = & \tilde{G}(i \omega) G(i\omega) U(i \omega) - \tilde{G}(i \omega) X(i \omega) \label{eq:band_limited_estimate} \\
    H(e^{i \Omega}) & = & \tilde{G}(i \omega) G(i\omega) \mathrm{DAC}(i \omega) \label{eq:band_limited_filter}
\end{IEEEeqnarray}
for $|\omega T_s| = |\Omega| < \pi $.

The \ac{CBADC} concept applied to a \ac{CTSDM} is merely a different perspective. 
As such, a conventional decimation filter $H(z)$, as in \Fig{fig:sdm_est}, has a corresponding $\tilde{G}(s)$ for a given \ac{CTSDM} and can thus be represented as in \Fig{fig:est_cbadc}. 
However, the \ac{CBADC} concept was originally not intended as an analysis tool for existing \ac{CTSDM} designs.
Instead, 
the ambition is to inspire new kinds of modulators as the interpretation of $G(s)$, $X(s)$, and $Q(z)$ are significantly different between the two estimation models in \Fig{fig:estimation_model}. 
Two converter examples, derived from the \ac{CBADC} perspective, will be given in \Sec{sec:leapfrog} and \Sec{sec:leap_frog_quadrature} respectively.

\section{The Leapfrog Analog Frontend}\label{sec:leapfrog}
In \Sec{sec:fundamentals}, we established that the foundation of the \ac{CBADC} concept is the continuous-time observation \Eq{eq:ct_observation}. Furthermore, we saw that a digital estimator, performing \Eq{eq:ct_estimate}, infers the \ac{CBADC} estimation model in \Fig{fig:est_cbadc}. As is evident from both \Eq{eq:ct_observation} and \Fig{fig:est_cbadc}, the fundamental conversion performance is determined by the open-loop gain corresponding to $g(t)$ and the largest magnitude of $|x(t)|$ respectively.
Therefore, a performant analog frontend is designed to have an analog system, i.e., $g(t)$, with a large open-loop gain within the frequency band of interest while simultaneously being stabilized by a digital control ensuring a bounded output swing on the final state $x(t)$.

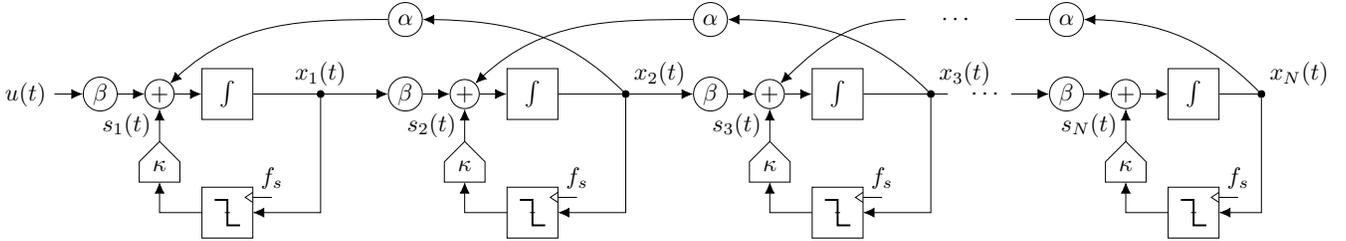
\begin{figure*}[tbp]
    \begin{center}
        \resizebox{\textwidth}{!}{
            \input{figures/leapfrog_baseband}
        }
        \caption{\label{fig:leapfrog-baseband}%
        The homogeneous leapfrog analog frontend.
        }
        
    \end{center}
\end{figure*}
Generally, we are not restricted to single-input-single-output loop-filters and a single digital feedback structure.
Instead, we can pursue a larger, more complex analog system that builds up frequency-selective amplification over several stages while ensuring stability by several local digital controls. The homogeneous low-pass leapfrog analog frontend \cite{M:20}, illustrated in \Fig{fig:leapfrog-baseband}, exemplifies this idea. Specifically, the leapfrog's analog system constitutes $N$ integrators that are interconnected in a leapfrog pattern by a forward gain $\beta$ and feedback gain $\alpha$. In addition, $N$ local digital controls ensure stability by feeding back 
the control signals $s_1(t),\dots,s_N(t)$. Each individual control signal is generated as in \Eq{eq:ct_control_signal} for a non-return to zero \ac{DAC}, i.e.,
\begin{IEEEeqnarray}{rCl}
    \mathrm{DAC}(t) & \eqdef & \kappa \cdot  \theta(t) \label{eq:dac_equation} \\
    \theta(t) & \eqdef & \begin{cases}1 & \mathrm{if} \, 0 < t \leq T_s \\ 0 & \mathrm{otherwise}\end{cases} \label{eq:non-return_impulse_response}
\end{IEEEeqnarray}
where $\kappa$ is an amplification factor.
Furthermore, each corresponding control decision $s_\ell[k]$ results from the output of a clocked comparator whose input is connected to the $\ell$th state variable $x_\ell(t)$.

Similarly to \Eq{eq:ct_observation}, from the \ac{CTSDM} example in \Sec{sec:fundamentals}, the leapfrog analog frontend's principle observation is the relation 
\begin{IEEEeqnarray}{rCl}
    x_N(t) & = & (g_u \ast u)(t) + \sum_{\ell=1}^N (g_{s_\ell} \ast s_\ell)(t) \label{eq:general_cbadc_observation}    
\end{IEEEeqnarray}
where $g_u$ is the impulse response from $u(t)$ to $x_N(t)$, in the absence of control signals, and $g_{s_\ell}(t)$ is the impulse response from $s_\ell(t)$ to $x_N(t)$ in the absence of input signal and all other control signals. 
Our interpretation of \Eq{eq:general_cbadc_observation} is that the joint digital control effort
, i.e., $\sum_{\ell=1}^N (g_{s_\ell} \ast s_\ell)(t)$, 
which is known via $g_{s_1}(t),\dots,g_{s_N}(t)$, \Eq{eq:dac_equation}, and $s_1[k], \dots, s_N[k]$, 
can be made an arbitrarily good continuous-time approximation of $(g_u \ast u)(t)$ 
for a relatively large analog system gain in relation to the maximum swing of $x_N(t)$.

\subsection{Parameterizing the Leap Frog analog frontend}\label{sec:leapfrog_parametrization}
Among the selling points of the leapfrog analog frontend in \Fig{fig:leapfrog-baseband}, is the ease of parametrization for a nominal target \ac{SNR} at a given bandwidth $\omega_\mathcal{B}$ [rad/s] while ensuring guaranteed stable operations \cite{FMWLY:2023}. Specifically,
\begin{IEEEeqnarray}{rCl}
    \mathrm{SNR} & = & \frac{g_i(N) \cdot \left(\frac{\mathrm{OSR}}{2\pi}\right)^{2N}}{\xi} \label{eq:SNR_prediction}
\end{IEEEeqnarray}
where $g_i(N) \approx 2^{2N - 1}$ can be computed analytically and $\xi$
is an empirical constant.
Furthermore, the specified performance, bandwidth,
and guaranteed stability follows from assigning $\alpha$, $\beta$, and $\kappa$ according to the relations
\begin{IEEEeqnarray}{rCl}
    |\beta| & = & \frac{\wb \cdot \text{OSR}}{2 \pi} \labell{eq:stability} \\
    \kappa & = & -\beta = \frac{\wb^2}{4\alpha}. \labell{eq:leapfrog_relation}
\end{IEEEeqnarray}

\cite{FMWLY:2023} showed that the design equations \Eq{eq:SNR_prediction}-\Eq{eq:leapfrog_relation} result in nominal performance similar to that of a heuristically optimized $\ac{CTSDM}$ with the same, \ac{OSR}, loop-filter order $N$, and an $N$ level quantizer.

\subsection{State-Space Equations}
In general
a \ac{CBADC} analog system is conveniently described using the underlying system of linear differential equations, i.e., state-space equations
\begin{IEEEeqnarray}{rCl}
    \dot{\vct{x}}(t) & = & \mat{A}_{\text{LP}} \vct{x}(t) + \mat{B}_{\text{LP}} u(t) + \vct{s}(t) \label{eq:state_equations_leapfrog} 
\end{IEEEeqnarray}
where
\begin{IEEEeqnarray}{rCl}
    \vct{x}(t) &\eqdef& \begin{pmatrix}x_1(t), \dots, x_N(t)\end{pmatrix}^\T, \\
    \vct{s}(t) &\eqdef& \begin{pmatrix}s_1(t), \dots, s_N(t)\end{pmatrix}^\T,
\end{IEEEeqnarray}
$\mat{A}_{\text{LP}} \in \R^{N\times N}$ and $\mat{B}_{\text{LP}} \in \R^N$. 

State space equations apply generally to filters and connect to a transfer function specification as  
\begin{IEEEeqnarray}{rCl}
    \begin{pmatrix}G_1(s), & \dots,& G_N(s) \end{pmatrix}^\T & = & \left(s \mat{I} - \mat{A}_{\text{LP}} \right)^{-1}\mat{B}_{\text{LP}}
\end{IEEEeqnarray}
where
\begin{IEEEeqnarray}{rCl}
    G_\ell(s) & \eqdef & \frac{X_\ell(s)}{U(s)}
\end{IEEEeqnarray}
is the transfer function from the filter input to any of its internal nodes, among which one is the filter output. Subsequently,
any \ac{CTSDM}'s loop filter, including the one from \Sec{sec:fundamentals}, can be characterized by an $A_{\mathrm{LP}}$ matrix and $B_{\mathrm{LP}}$ vector.
In particular, the leapfrog analog system, from \Fig{fig:leapfrog-baseband}, 
reduces to the state space matrices
\begin{IEEEeqnarray}{rCl}
    \mat{A}_{\text{LP}} & \eqdef &
    \begin{pmatrix}
        0 & \alpha \\
        \beta & 0 & \ddots \\ 
         & \ddots & \ddots & \alpha \\
         &  & \beta & 0  \\
    \end{pmatrix} \\ 
    \mat{B}_{\text{LP}} & \eqdef & \begin{pmatrix}\beta, 0,\dots, 0\end{pmatrix}^\T.
\end{IEEEeqnarray}

\section{The CBADC Digital Estimator}\label{sec:general_digital_estimator}
Generalizing the special case of a single control signal, as in the \ac{CTSDM} from \Sec{sec:cbadc_est}, the more general \ac{CBADC} digital estimator
performs the discrete-time convolution
\begin{IEEEeqnarray}{rCl}
    \hat{u}[k]  & \eqdef & \sum_{\ell = 1}^N \sum_{k_1 \in \Z} h_\ell[k_1] s_\ell[k - k_1] \label{eq:estimate}
\end{IEEEeqnarray}
which once more, using the \ac{CBADC} principal observation \Eq{eq:general_cbadc_observation}, can be seen as the sampling operation
\begin{IEEEeqnarray}{rCl}
    \hat{u}[k] & \eqdef & (\tilde{g} \ast g_u \ast u)(k T_s) + (\tilde{g} \ast x_N)(k T_s) \label{eq:sampling}
\end{IEEEeqnarray}
for 
\begin{IEEEeqnarray}{rCl}
    h_\ell[k] & \eqdef & \kappa (\tilde{g} \ast g_{s_\ell} \ast \theta)(k T_s). \label{eq:discrete_filter_definition}
\end{IEEEeqnarray}
The steps involved in going from \Eq{eq:estimate} to \Eq{eq:sampling} are worked out in \App{app:sampling_decomposition_multi_controls}. 
An immediate consequence of \Eq{eq:sampling} is that the estimation model in \Fig{fig:est_cbadc},
assuming $x_N(t)$ to be an independent variable, 
also applies for a general analog frontend with a digital estimator implementing \Eq{eq:estimate} as the \ac{CTSDM}'s loop-filter $G(s)$ and $X(s)$ are replaced by the Laplace transform of analog system's impulse response $g_u(t)$ and $x_N(t)$ in \Fig{fig:est_cbadc} respectively.

\subsection{Adaptive-Filter Calibration}\label{sec:calibration}
In contrast to the scalar case, 
\Eq{eq:discrete_filter_definition} introduces a new problem;
namely, fixing a single $h_\ell[k]$ in \Eq{eq:estimate} not only determines $\tilde{g}(t)$ but also all other $h_{k \neq \ell}[k]$.
This makes the computation of $h_{k \neq \ell}[k]$ most dependent on analog system knowledge and the general estimator sensitive to variations thereof \cite{FMWLY:2022, FMWLY:2023}.

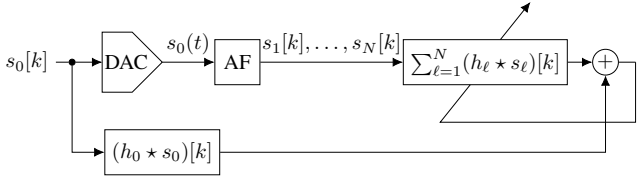
\begin{figure}
    \centering
    \resizebox{0.99\columnwidth}{!}{
        \input{figures/adaptive_calibration}
    }
    \caption{\label{fig:adaptive_calibration}%
    The adaptive-filter calibration which minimizes \Eq{eq:adaptive_optimization_problem} to estimate the filter coefficients in \Eq{eq:estimate} where $(h_\ell \star s_\ell)[k] \eqdef \sum_{k_1\in\Z} h_\ell[k_1]s_\ell[k-k_1]$ refers to discrete-time convolution and the analog frontend (AF) generates the calibration sequences.
    }
\end{figure}

The authors in \cite{MMBFL:22} showed that the filter coefficients do not need to be computed analytically.
Instead, these can be found via a foreground calibration where $u(t)$ is replaced by a broadband reference control signal $s_0[k]$ during calibration. 
The calibration process is an adaptive filtering problem \cite{Haykin:2002} which is outlined in \Fig{fig:adaptive_calibration} and amounts to solving the convex optimization problem
\begin{IEEEeqnarray}{rCl}
    \argmin_{h_1,\dots,h_N} \sum_{k} \left|\sum_{\ell = 0}^N \sum_{k_1 \in \Z} h_\ell[k_1] s_\ell[k - k_1] \right|^2 \label{eq:adaptive_optimization_problem}
\end{IEEEeqnarray}
where 
\begin{IEEEeqnarray}{rCl}
    h_0[k] & \eqdef & -\kappa (\tilde{g} \ast g_u \ast \theta)(k T_s) 
\end{IEEEeqnarray}
is a fixed reference filter and $h_1[k], \dots, h_N[k]$ are determined through calibration.
We stress that choosing $h_0[k]$ via conventional discrete-time filter design techniques and solving \Eq{eq:adaptive_optimization_problem} are the practical tasks of the digital estimator, whereas $\tilde{g}(t)$ and the sampling perspective in \Eq{eq:sampling} follows implicitly upon a successful calibration
and the analog frontend parameterization. Furthermore, when fixing $h_0[k]$, such that $\tilde{G}(s)$ band-limits both terms in \Eq{eq:sampling}, follows \Eq{eq:band_limited_estimate} and \Eq{eq:band_limited_filter} for $G(s) = G_u(s)$ since both the reference sequence $s_0(t)$ and $u(t)$ have the same transfer function to $x_N(t)$.

\subsubsection{Adaptive LMS Filtering}\label{sec:lms_calibration}
One way of approaching the calibration problem in \Eq{eq:adaptive_optimization_problem} is the classical \ac{LMS} adaptive filter \cite{Haykin:2002}. For an \ac{LMS} adaptive filter, we assume $h_1[k],\dots,h_N[k]$ to be \ac{FIR} filters. Subsequently, the gradient of \Eq{eq:adaptive_optimization_problem} with respect to the filter weight $h_\ell[m]$ follows as 
\begin{IEEEeqnarray}{rCl}
    2 \left(\hat{u}[k] + \sum_{k_1} h_0[k_1]s_0[k-k_1]\right) s_\ell[k - m] \label{eq:lms_gradient}
\end{IEEEeqnarray}
Remarkably, in the case of binary control signals $s_0[k], \dots, s_N[k]$, i.e., $s_\ell[m]\in\{\pm1\}$,
the estimator in \Eq{eq:estimate} and the gradient step from \Eq{eq:lms_gradient} can be computed without multiplications, suggesting that
the sampling perspective in \Eq{eq:sampling}
can be enforced using only digital additions.

\subsection{The Control-Bounded A/D Design Perspective}

The digital estimator is the enabler of the control-bounded approach to \ac{A/D} conversion.
However, 
rather than its operation, 
it is the design perspective that it facilitates that is the essence of this approach.
The control-bounded design perspective springs from the three following key observations
\begin{enumerate}
\item The analog frontend design determines the nominal conversion performance \Eq{eq:general_cbadc_observation}.

\item The digital estimator's operation, i.e.,  \Eq{eq:estimate} and \Eq{eq:adaptive_optimization_problem}, only involves digital discrete-time operations on the $s_0[k],\dots,s_N[k]$ signals and is otherwise indifferent to the analog frontend design.

\item Calibration bridges analog and digital by enforcing the sampling perspective in \Eq{eq:sampling}.

\end{enumerate}

\section{Quadrature Analog Frontends} \label{sec:quadrature}
Any low-pass \ac{CBADC} analog systems, for example, the leapfrog analog system from \Sec{sec:leapfrog} or the \ac{CTSDM} analog system from \Sec{sec:fundamentals}, can be extended into a corresponding \ac{QCBADC}. 
The \ac{QCBADC} follows 
from two modifications:
Firstly, interconnecting two low-pass analog systems such that they oscillate at the desired notch frequency $f_n$, further described in \Sec{sec:qas}.
Secondly, stabilizing the resulting analog system using a local quadrature digital control, which follows in \Sec{sec:lqdc}. 

\subsection{Quadrature Analog System}\label{sec:qas}
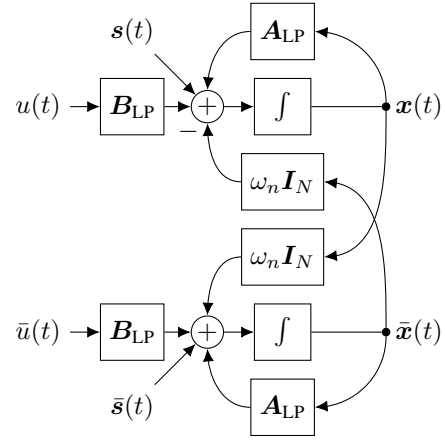
\begin{figure}
    \centering
        \input{figures/quadrature_extension}
    \caption{\label{fig:quadrature_analog_system}%
        A general quadrature analog system where two instances of a low-pass analog system, $\mat{A}_{\text{LP}}$ and $\mat{B}_{\text{LP}}$, are connected such that the resulting transfer function from $(u(t), \bar{u}(t))$ to $(x_N(t), \bar{x}_N(t))$ is centered around the angular notch frequency $\omega_n$.
    }
\end{figure}

For a quadrature analog system to oscillate at a desired angular frequency $\omega_n = 2 \pi f_n$,
two identical $N$th order analog systems are stacked in parallel and interconnected as
\begin{IEEEeqnarray}{rCl}%
    \begin{pmatrix}\dot{\vct{x}}(t) \\ \dot{\bar{\vct{x}}}(t)\end{pmatrix} & = & \mat{A} \begin{pmatrix}\vct{x}(t) \\ \bar{\vct{x}}(t)\end{pmatrix} + \mat{B}\begin{pmatrix}u(t) \\ \bar{u}(t) \end{pmatrix} + \begin{pmatrix}\vct{s}(t) \\ \bar{\vct{s}}(t)\end{pmatrix}  \labell{eq:state_space_equation} \\
    \mat{A} & \eqdef & \begin{pmatrix} \mat{A}_{\text{LP}} & -\omega_n \mat{I}_N \\ \omega_n \mat{I}_N &  \mat{A}_{\text{LP}} \end{pmatrix} \in \R^{2 N \times 2 N} \label{eq:quadrature_A} \\
    \mat{B} & \eqdef & \begin{pmatrix} \mat{B}_{\text{LP}} & \mat{0}_N \\ \mat{0}_N & \mat{B}_{\text{LP}} \end{pmatrix} \in \R^{2 N \times 2}  \label{eq:ssm_quad_3}
\end{IEEEeqnarray}
where $\mat{A}_{\text{LP}}$ and $\mat{B}_{\text{LP}}$ refers to the system description of a low-pass analog system, e.g., the $\mat{A}_{\text{LP}}$ and $\mat{B}_{\text{LP}}$ from \Sec{sec:leapfrog}, $\mat{I}_N \in \R^{N \times N}$ is an identity matrix, and 
\begin{IEEEeqnarray}{rCl}
    \vct{x}(t) & = & \begin{pmatrix}x_1(t),\dots,x_N(t)\end{pmatrix}^\T \\
    \bar{\vct{x}}(t) & = & \begin{pmatrix}\bar{x}_1(t),\dots,\bar{x}_N(t)\end{pmatrix}^\T \\
    \vct{s}(t) & = & \begin{pmatrix}s_1(t),\dots,s_N(t)\end{pmatrix}^\T \\
    \bar{\vct{s}}(t) & = & \begin{pmatrix}\bar{s}_1(t),\dots,\bar{s}_N(t)\end{pmatrix}^\T 
\end{IEEEeqnarray}
together with $u(t)$ and $\bar{u}(t)$ are the in-phase and quadrature parts of the state vector, control signal vector, and input signal, respectively.
\Fig{fig:quadrature_analog_system}, visualize the meaning of \Eq{eq:state_space_equation}-\Eq{eq:ssm_quad_3} as a low-pass \ac{CBADC} analog system is transformed into a corresponding quadrature version.

\subsection{Local Quadrature Digital Control}\label{sec:lqdc}
\begin{figure*}[tbp]
    \begin{center}
            \input{figures/quadrature_dc}
        \caption{\label{fig:digital_control_zero_order_hold}%
        The $\ell$th local quadrature digital control, stabilizing the quadrature state pair $\begin{pmatrix}x_\ell(t), \bar{x}_\ell(t)\end{pmatrix}^\T$ using the quadrature control signal pair $\begin{pmatrix}s_\ell(.),\bar{s}_\ell(.)\end{pmatrix}^\T$.
        }
    \end{center}
\end{figure*}
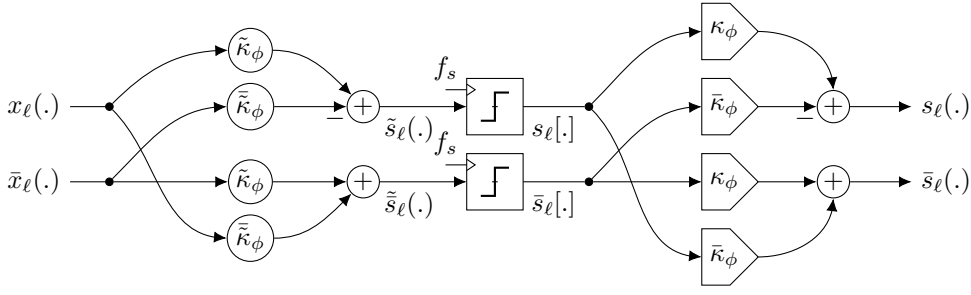
A quadrature analog system, as in \Fig{fig:quadrature_analog_system}, 
can be stabilized by $N$ local quadrature digital controls as shown in \Fig{fig:digital_control_zero_order_hold}. 
The $\ell$th digital control stabilizes the quadrature state pair $\begin{pmatrix}x_\ell(t), \bar{x}_\ell(t)\end{pmatrix}^\T$
via the quadrature control observations 
\begin{IEEEeqnarray}{rCl}%
    \begin{pmatrix}\tilde{s}_\ell(t) \\  \bar{\tilde{s}}_\ell(t)\end{pmatrix} & \eqdef & \begin{pmatrix}\tilde{\kappa}_\phi & - \bar{\tilde{\kappa}}_\phi \\ \bar{\tilde{\kappa}}_\phi  & \tilde{\kappa}_\phi\end{pmatrix} \begin{pmatrix}x_\ell(t) \\ \bar{x}_\ell(t)\end{pmatrix},\label{eq:control_observation}
\end{IEEEeqnarray}
the quadrature control contribution
\begin{IEEEeqnarray}{rCl}    
    \begin{pmatrix}s_\ell(t) \\ \bar{s}_\ell(t)\end{pmatrix} 
    & \eqdef & \sum_{k \in \Z}  \begin{pmatrix}\kappa_\phi & -\bar{\kappa}_\phi \\ \bar{\kappa}_\phi & \kappa_\phi\end{pmatrix} \begin{pmatrix}\theta(t - \tau_{\text{DC}} - kT_s)  s_\ell[k] \\ \theta(t - \tau_{\text{DC}} - kT_s)  \bar{s}_\ell[k]\end{pmatrix}, \nonumber \\  \label{eq:control_signal}
\end{IEEEeqnarray}
and the quadrature control signal $\begin{pmatrix}s_\ell[k], \bar{s}_\ell[k]\end{pmatrix}^\T$.
As for the regular local digital control, the control signals are generated with a non-return to zero \ac{DAC}, cf. \Eq{eq:non-return_impulse_response} where additionally we introduce $\tau_{\text{DC}}>0$, the time delay associated with the sampling and quantization process.

\subsection{Parameterize for Guaranteed Stability}
Next, we seek values of $\kappa_\phi$, $\bar{\kappa}_\phi$, $\tilde{\kappa}_\phi$, $\tilde{\bar{\kappa}}_\phi$, and $f_s$, such that $N$ local quadrature digital controls as in \Fig{fig:digital_control_zero_order_hold} stabilizes a given analog system as in \Fig{fig:quadrature_analog_system}.

Following the approach in \cite{M:20}, stability can be ensured if each pair-wise quadrature state system can be stabilized independently for a bounded worst-case input signal with a local quadrature digital control. Specifically, the global system stability follows as the input to each pair-wise quadrature system can only combine other bounded pair-wise states and the bounded input signal.

Assuming no quadrature state pair or input pair feeds coherently into another quadrature state pair with a combined amplification larger than $\beta$,
the parametrization 
\begin{IEEEeqnarray}{rCl}%
    \kappa_\phi & = &  \frac{\beta T_s \omega_n}{2 \sin\left(\frac{\omega_n T_s}{2}\right)}\cos\left(\phi_\kappa\right) \label{eq:kappa_phi}\\
    \bar{\kappa}_\phi & = &  \frac{\beta T_s \omega_n}{2 \sin\left(\frac{\omega_n T_s}{2}\right)} \sin\left(\phi_\kappa\right) \\
    \tilde{\kappa}_\phi & = & -\frac{1}{\beta T_s}\cos\left(\omega_n \left(\frac{T_s}{2} + \tau_{\text{DC}}\right) - \phi_\kappa\right)\\
    \bar{\tilde{\kappa}}_\phi & = & -\frac{1}{\beta T_s}\sin\left(\omega_n \left(\frac{T_s}{2} + \tau_{\text{DC}}\right) - \phi_\kappa\right) \label{eq:bar_tilde_kappa_phi} \\
    f_s & = & 2 \beta \label{eq:control_rate}
\end{IEEEeqnarray}
guarantees a bounded state pair where $\phi_\kappa\in[0, 2\pi)$ is a free parameter that may be chosen to ensure practical values.
A detailed derivation of \Eq{eq:kappa_phi}-\Eq{eq:control_rate} is given in \App{app:lqdc}.
The abstract meaning of $\beta$, in \Eq{eq:kappa_phi}-\Eq{eq:control_rate}, will be made concrete with the quadrature leapfrog analog frontend in \Sec{sec:leap_frog_quadrature}. 

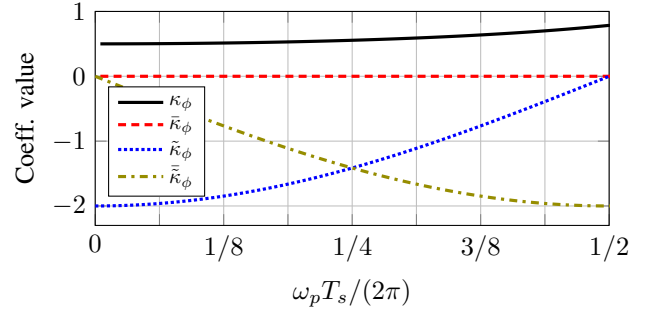
\begin{figure}
    \centering
    \input{figures/kappa_figure}
    \caption{\label{fig:kappa_figure} The coefficients in \Eq{eq:kappa_phi}-\Eq{eq:bar_tilde_kappa_phi} as a function of $\omega_n T$ where $2 \beta T_s = 1$, $\phi_\kappa = 0$, and $\tau_{DC} = 0$.}
\end{figure}

\Fig{fig:kappa_figure} demonstrates how $\kappa_\phi$, $\bar{\kappa}_\phi$, $ \tilde{\kappa}_\phi$, and $\tilde{\bar{\kappa}}_\phi$ depends on $\omega_n T_s$ for $2 \beta T_s = 1$, $\phi_\kappa = 0$, and $\tau_{DC} = 0$.

\subsection{On Frequency Interleaving}
We can imagine application scenarios where several \acp{CBADC}, or \acp{BPSDM}, are used to increase the combined \ac{ADC}'s signal bandwidth \cite{LF:2023}.
Alternatively, converting input signals that spans several disjoint frequency bands.

In both these scenarios, the \ac{QCBADC} ability to parameterize for an arbitrary notch frequency makes it an attractive building block.
Perhaps less obvious, for multiple \acp{QCBADC} scenarios, the calibration, outlined in \Sec{sec:calibration}, 
applied jointly to all control signals would produce a unified estimate.
Furthermore, the computational effort scales linearly with the number of control signals for adaptive \ac{LMS} filtering, see \Sec{sec:lms_calibration}.
Therefore, calibrating several digital estimators separately or jointly comes asymptotically at the same computational cost.
Unfortunately, interleaved \ac{QCBADC} structures fall outside the scope of this paper.

\section{The Leapfrog Quadrature Analog Frontend} \label{sec:leap_frog_quadrature}
\begin{figure*}
    \centering
    \resizebox{\textwidth}{!}{    
        \input{figures/quadrature_full}
    }
    \caption{\label{fig:quadrature_simplified}%
    The leapfrog quadrature analog frontend for a local quadrature digital control as in \Fig{fig:digital_control_zero_order_hold} where $\phi_\kappa = 0$.
    }
\end{figure*}
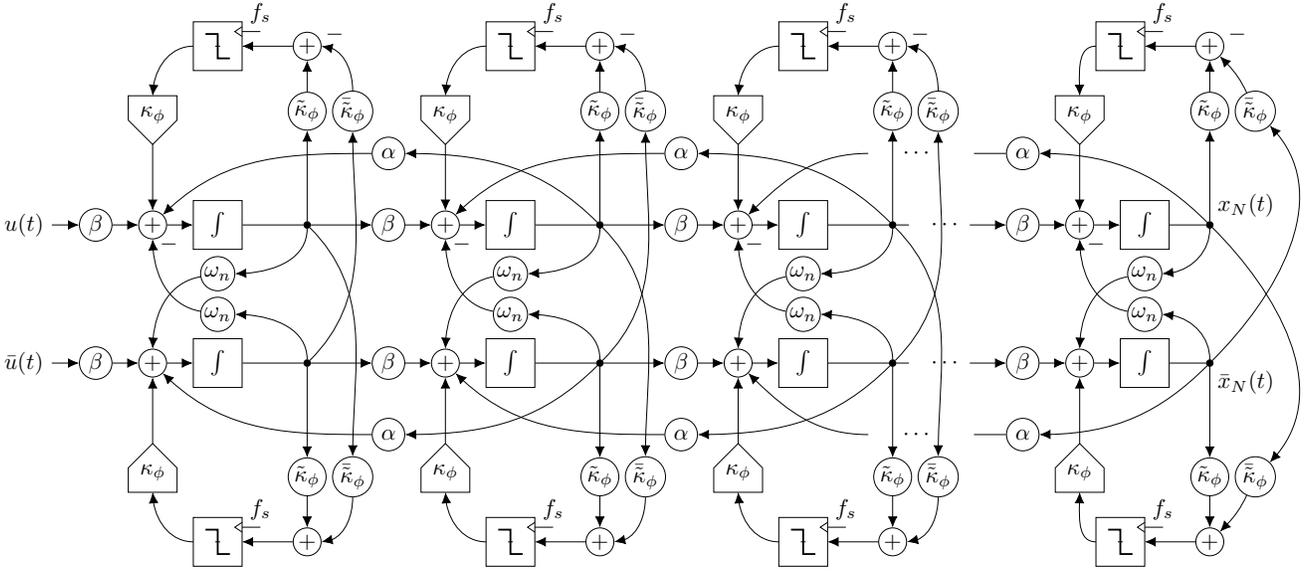
\begin{figure}
    \centering
    \input{figures/tf_lfs}
    \caption{\label{fig:lf_tf}%
    Transfer function, from $u(t)$ to $x_N(t)$, for the leapfrog analog system in \Fig{fig:quadrature_simplified} where $\mathrm{OSR}=4$ and $X_N(i2\pi f)$ and $U(i2 \pi f)$ correspond to the Laplace transforms of $x_N(t)$ and $u(t)$ respectively.
    }
\end{figure}
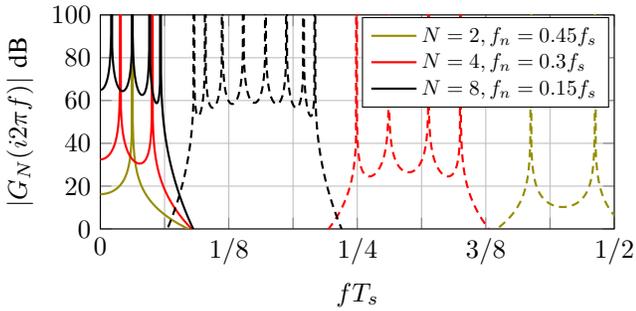

The quadrature version of the leapfrog analog frontend from \Sec{sec:leapfrog} is shown in \Fig{fig:quadrature_simplified} for $\phi_\kappa = 0$.
Although its complex appearance, we recognize the fundamental building blocks, i.e., two low-pass leapfrog analog system structures from \Fig{fig:leapfrog-baseband}, $N$ local quadrature digital controls as in \Fig{fig:digital_control_zero_order_hold}, and the multiplicative interconnects $\omega_n$.
As in the case of a low-pass leapfrog analog system, the modular structure can be seamlessly adapted by changing the number of stages, i.e., $N$, where a larger $N$
results in a larger \ac{SNR} or alternatively a lower \ac{OSR} for the same \ac{SNR}.
We consider the analog system transfer function for both structures in \Fig{fig:lf_tf}.
Specifically, for an $\mathrm{OSR}=4$, the low-pass transfer function is plotted in solids and the corresponding quadrature version in dashed
for three different system orders $N=(2,4,8)$ where the notch frequency is chosen as $f_n=(0.45 f_s, 0.3 f_s, 0.15 f_s)$ respectively.

A perhaps alarming visual from \Fig{fig:lf_tf} is the presence of undamped poles in the transfer function. 
To this end, we remind ourselves that the analog system is accompanied by a digital control designed to guarantee stable operations, cf. \Sec{sec:lqdc}. 
Furthermore, as per the control-bounded observation \Eq{eq:general_cbadc_observation}, for a digital control which bounds $x_N(t)$ by the same bound for any $N$,
the conversion performance follows directly from the open-loop gain, from $u(t)$ to $x_N(t)$, in the transfer function \Fig{fig:lf_tf}. In other words, 
for a structure as the low-pass and quadrature leapfrog analog frontend in \Fig{fig:leapfrog-baseband} and \Fig{fig:quadrature_simplified}, a larger $N$ means a larger open-loop gain and thereby a better A/D conversion performance.

From comparing the low-pass and quadrature transfer functions in \Fig{fig:lf_tf}, we also recognize that the latter has twice the bandwidth and number of poles compared to the former. This is expected as we now have twice the number of integrators. 
Additionally, twice the bandwidth follows as the input signal is a quadrature signal $(u(t), \bar{u}(t))$, and the amplitude response from $\bar{u}(t)$ to $\bar{x}_N(t)$ is identical to the amplitude response from $u(t)$ to $\bar{x}_N(t)$ given in \Fig{fig:lf_tf}. Importantly, this does not imply that we have twice the bandwidth for two independent signals $u(t)$ and $\bar{u}(t)$.

\subsection{Behavioural Simulations}\label{sec:simulation}

Behavioral simulations of the low-pass and quadrature leapfrog, in \Fig{fig:leapfrog-baseband} and \Fig{fig:quadrature_simplified}, combined with a corresponding digital estimator, were done in Python using the cbadc toolbox \cite{cbadc:2022}.
Each simulation used a full-scale input signal $u(t) = v_{\text{fs}} \cos(2 \pi f_{\mathrm{LP}} t)$ alternatively a quadrature pair of full-scale input signals,
$u(t) = v_{\text{fs}} \cos(2 \pi f_{\mathrm{BP}} t)$ and $\bar{u}(t) = v_{\text{fs}} \sin(2 \pi f_{\mathrm{BP}} t)$ for $f_{\mathrm{LP}} \eqdef \omega_{\mathcal{B}} / (4 \pi)$ and $f_{\mathrm{BP}} \eqdef f_n - \omega_{\mathcal{B}} / (8 \pi)$ respectively.
Additionally, the sampling and quantization process is assumed to be without delay, i.e.,
$\tau_{\text{DC}} = 0$.

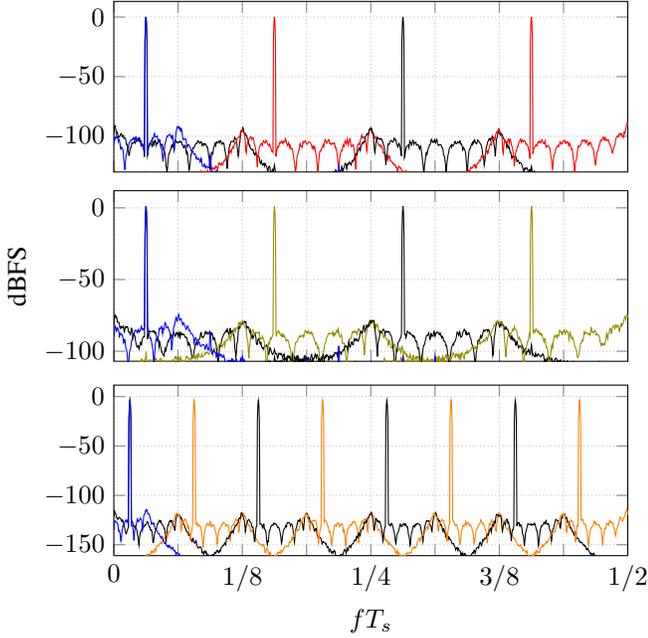
\begin{figure}
    \centering
    \input{figures/psd}
    \caption{\label{fig:psd} 
    \ac{PSD} for different notch frequencies. The black, red, green, and yellow lines correspond to \acp{PSD} of \Eq{eq:estimate} from \acp{QCBADC} designed for $\text{\ac{OSR}}/N=(4/8, 4/6, 8/6)$ (top, middle, bottom) and positioned at different notch filter frequencies $f_n$. 
    Similarly, the blue lines are the \acp{PSD} of a corresponding low-pass leapfrog building block. 
    }
\end{figure}

Several \ac{OSR}, $N$, and notch frequencies $f_n$ were considered to showcase the modular structure and flexible notch frequency.
The simulation results are visualized in \Fig{fig:psd} by plotting the \ac{PSD} of the final estimate, i.e., the output of the digital estimator, \Eq{eq:estimate}.
The \ac{PSD} is computed using a $2^{14}$ point FFT and normalized for \ac{dBFS}.
From the \Fig{fig:psd} we see three different configurations $\text{\ac{OSR}}/N=(4/8, 4/6, 8/6)$ (top, middle, bottom) where each plot contains $\mathrm{OSR} + 1$ estimated \acp{PSD} where the blue spectrum corresponds to the low-pass leapfrog building block and the rest are quadrature leapfrog analog frontends positioned to cover the whole $f_s / 2$ frequency range jointly.
The \acp{SNR}, measured directly on the \acp{PSD} in \Fig{fig:psd}, are approximately $83$, $67$, and $105$ dB (top, middle, bottom) with variations within $\pm 1$ dB between the different notch frequencies (including the low-pass case).

As these are nominal simulations, the digital estimator filter, \Eq{eq:estimate}, was not computed using calibration but analytically using the Wiener filter solution from \cite{MWL:21}. The Wiener filter implies the inverse filter
\begin{IEEEeqnarray}{rCl}
    \tilde{G}(s) & = & \frac{\bar{G}_u(s)}{\left|G_u(s)\right|^2 + \eta^2}
\end{IEEEeqnarray}
where $\bar{G}_u(s)$ refers to the complex conjugate of $G_u(s)$ and the bandwidth parameter was set as $\eta^2 = |G_u(i \omega_{\mathcal{B}})|^2$. 
Subsequently, 
assuming $x(t)$ is an independent quantity and that $\tilde{G}(s)$ effectively suppress all out-of-band signal, 
the transfer function from the input signal to the digital estimator's final estimate is then shaped by the \ac{STF}
\begin{IEEEeqnarray}{rCl}
    \frac{\hat{U}(e^{i \Omega})}{U(i\omega)} & = & \frac{\left|G_u(i\omega)\right|^2}{\left|G_u(i\omega)\right|^2 + |G_u(i \omega_{\mathcal{B}})|^2}
\end{IEEEeqnarray}
and the conversion error $x(t)$ will appear in the estimate via the \ac{NTF}
\begin{IEEEeqnarray}{rCl} 
    \frac{\hat{U}(e^{i \Omega})}{X(i\omega)} & = & \frac{\bar{G}_u(i\omega)}{\left|G_u(i\omega)\right|^2 + |G_u(i \omega_{\mathcal{B}})|^2}
\end{IEEEeqnarray}
for $|\omega T_s| = |\Omega| < \pi $ as per \Eq{eq:band_limited_estimate}. The Wiener filters $h_\ell[k]$ were implemented as truncated \ac{FIR} filters, each with $2^{12}$ taps. 
The number of taps was intentionally selected to be much larger than necessary to prevent numerical artifacts relating to the digital estimator.

\subsection{Mismatch Sensitivity}
A significant limiting factor for both \ac{BPSDM} and \ac{QSDM} is stability in the presence of component variations while targeting a wide tunable range \cite{AKYO:09,AL:04}.
The proposed \ac{QCBADC}'s sensitivity to component variations was tested by $256$ Monte Carlo simulations where each individual $(\alpha,\beta, \omega_n, \kappa_\phi,
\tilde{\kappa}_\phi,\tilde{\bar{\kappa}}_\phi)$ value
were drawn uniformly at random from within $\pm10$\% of their nominal values. 
For these simulations, the system was modeled using Verilog-A and simulated in Cadence Spectre with $\text{\ac{OSR}}=8, N=6$ and $f_n = f_s / 8$, which corresponded to $105$ dB nominal SNR, see \Fig{fig:psd} (bottom). 
The filter coefficients in \Eq{eq:estimate} were calculated analytically, as in \Sec{sec:simulation}, from the actual system parametrization corresponding to a perfectly calibrated digital estimator. 
The same component variation scenario was also simulated for the low-pass leapfrog building block.

\begin{figure}
    \centering
    \input{figures/hist_FS}
    \caption{\label{fig:hist} Histograms showing variation in \ac{SNR} relative to the nominal performance of $105$ dB (top) and estimated $\hat{f}_n$ normalized to nominal $f_n=f_s / 8$ (bottom) after 256 Monte Carlo simulations with up to $\pm 10\%$ variations in each $(\alpha,\beta,\omega_n,\kappa_\phi,\bar{\kappa}_\phi,\tilde{\kappa}_\phi,\tilde{\bar{\kappa}}_\phi)$ parameter.}
\end{figure}
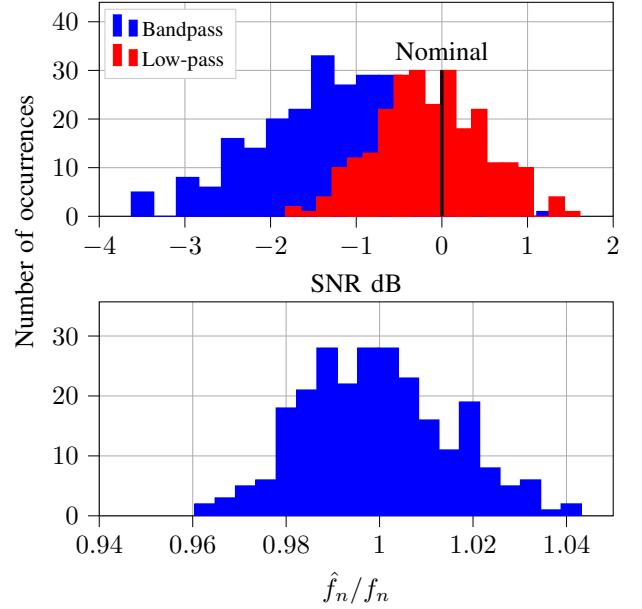

The simulated component variations resulted in $3/256$ of the leapfrog low-pass realizations being unstable, and for those realizations, a significant \ac{SNR} deficit.
In contrast, none of the quadrature leapfrog realizations resulted in any instability. Excluding the unstable low-pass leapfrog realizations, 
histograms of the resulting \ac{SNR} performance and estimated notch frequencies $\hat{f}_n$
are shown in \Fig{fig:hist}
The results show \acp{SNR} and notch frequencies $\hat{f}_n$ in the ranges of $(-4,2)$~dB, and $\pm 5$\% from their respective nominal values.  

We conjecture that
the average reduction of $\approx1$~dB \ac{SNR}, in the quadrature leapfrog case, originates from the fact that non-aligning notch frequencies per stage results in an open-loop gain loss in the transfer function from $(u(t), \bar{u}(t))$ to $(x_N(t), \bar{x}_N(t))$. 
In contrast, the per-stage bandwidth variations, which apply to both quadrature and low-pass leapfrog versions, 
appears to have no significant impact on the average \ac{SNR} based on the performance of the low-pass leapfrog version.

In contrast to the known instability sensitivities of \ac{BPSDM} and \ac{QSDM}, our findings suggest that the quadrature leapfrog analog frontend is not particularly sensitive to component variations. Perhaps surprisingly, the \ac{QCBADC} appears at least as robust as its leapfrog building block. 

\subsection{Op-amp Non-Idealities}
\begin{figure}[h]
    \begin{center}
            \input{figures/quadrature_implementation}
        \caption{\label{fig:quadrature_impl}%
        A single-ended op-amp implementation of the $\ell$th quadrature stage in \Fig{fig:quadrature_simplified}
        where the two $f_s$ clocked comparators implements a joint sampling, $\tilde{\kappa}$, $\bar{\tilde{\kappa}}$ weighted 1-bit quantization, and a subsequent \ac{DAC} stage. The digital control signals $s_\ell[k]$ and $\bar{s}_\ell[k]$ are accessible within these comparators.
        }
    \end{center}
\end{figure}
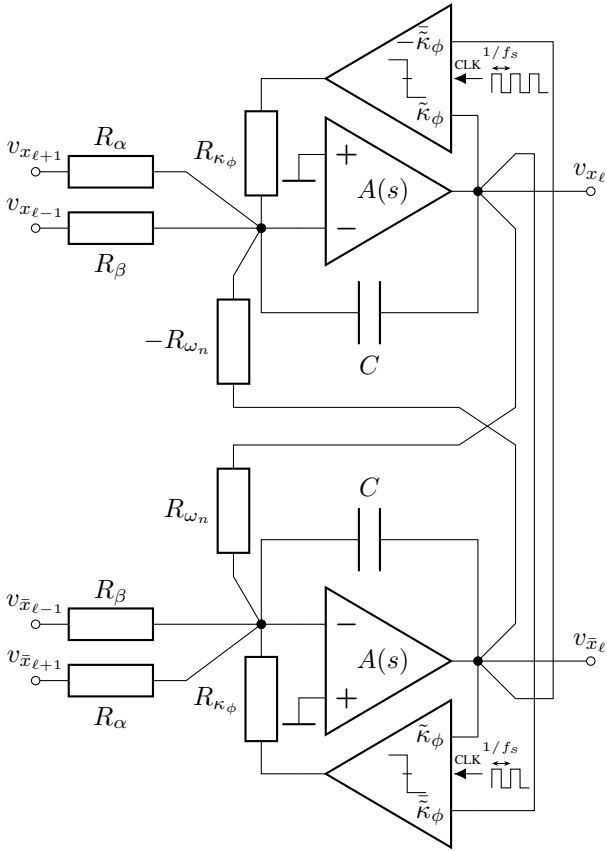

To further validate the robustness properties of the quadrature leapfrog analog frontend, we proceed by including first-order op-amp non-idealities.
Specifically, we consider implementing \Fig{fig:quadrature_simplified} using an inverting op-amp circuit topology.
The $\ell$th quadrature pair would then be realized as shown in \Fig{fig:quadrature_impl} where the two op-amps are characterized by their transfer function $A(s)$, 
the integrators are realized with capacitive feedback $C$, 
and all interconnects follows from the resistors as $R_{\alpha}C = \alpha^{-1}$, $R_{\beta}C = \beta^{-1}$, $R_{\kappa}C = \kappa^{-1}$, and $R_{\omega_n}C = \omega_n^{-1}$. The local quadrature digital control is implemented using two clocked comparators that effectively quantize and sample a $(\tilde{\kappa}_\phi,\bar{\tilde{\kappa}}_\phi)$ weighted snapshot of the quadrature state pair $(v_{x_\ell}, v_{\bar{x}_\ell})$ and realizes the non-return to zero \ac{DAC} by outputting the resulting control signals via the two $R_{\kappa_\phi}$.

The circuit is parameterized with a $f_s = 2^{31} \approx 2.1$ GHz, $N=6$, $\mathrm{OSR}=4$, and $f_n = 5/16 f_s \approx 671 $ MHz resulting in $\omega_{\mathcal{B}} \approx 268 $ $\mathrm{MHz}/(2 \pi)$, and a target \ac{SNR} performance of $65$ dB. For $C=1$ pF, it follows that $R_\beta \approx 931$ $\Omega$, $R_\alpha \approx 6.04$ k$\Omega$, $R_\kappa \approx 788$ $\Omega$, and $R_{\omega_n} \approx 237$ $\Omega$.
These component values are chosen without much afterthought and are perhaps problematic for a given technology. However, the $RC$ products are the relevant quantity for this circuit simulation.

The first-order op-amp non-idealities are enforced by introducing a finite DC-gain $k_A$ and \ac{GBWP} $k_A \omega_A$  via the op-amp transfer function 
\begin{IEEEeqnarray}{rCl}
    A(s) & = & k_{A} \frac{\omega_{A}}{s + \omega_{A}}. \label{eq:op-amp-first-order}
\end{IEEEeqnarray}

Practically, the introduction of an internal op-amp pole in \Eq{eq:op-amp-first-order} makes calibration of the digital estimator necessary as we then implement a different system of differential equations from \Fig{fig:quadrature_simplified}. 
Although it is possible to compute the corresponding Wiener filter for the extended system of differential equations, relying on precise \ac{GBWP} and DC-gain values in a practical realization is not realistic. 

Transient simulations were done using the open-source Ngspice software, where automated netlists were generated using the cbadc toolbox \cite{cbadc:2022} and the calibration from \Sec{sec:calibration} was applied to the resulting control signals.
This means that each configuration was simulated both with an input signal (testing) and without an input signal (training), and calibration was performed on the latter, whereas the resulting \acp{PSD} were computed on the former. 

\begin{figure}
    \centering
    \input{figures/calibrated_filter}
    \caption{\label{fig:calibrated_filters}%
    The calibrated $h_0[k],\dots,h_{N}[k]$ filter's amplitude response for $k_A \omega_A (f_n + \frac{\omega_\mathcal{B}}{4 \pi}) = 750$ and $\kappa_A = 10^4$. Note that $\bar{h}_{0}[k], \dots, \bar{h}_{N}[k]$ have identical amplitude responses to their corresponding in-phase filter.
    }
\end{figure}
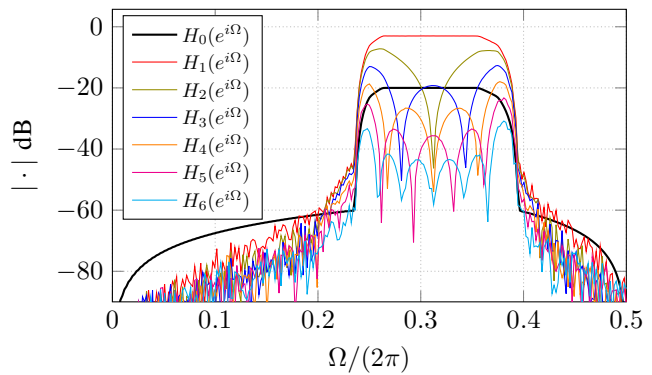

The in-phase and quadrature digital estimation filters $h_0[k],\dots,h_N[k]$ and $\bar{h}_0[k],\dots,\bar{h}_N[k]$ were all chosen as \ac{FIR} filters with $2^9$ taps, and initialized with all zero coefficients except for $h_0[k] = \bar{h}_0[k]$ which was designed with standard \ac{FIR} design techniques to have an amplitude response of $(0, -3, -20, -\infty, -\infty)$ dB at 
$(671, 671 \pm 134, 671 \pm 141, 0, 1074)$ MHz, shown in the amplitude response $H_0(e^{i\Omega})$ in \Fig{fig:calibrated_filters}. 
Furthermore, two randomly generated binary reference sequence $s_0[k]$, $\bar{s}_0[k]$ entered the first quadrature pair via a $R_{\kappa_{0}} = 10 R_{\kappa_\phi}$. This allowed the reference sequence to be active for training and testing conditioned on a reduced input signal of $9/10 v_{\text{fs}}$ where $v_{\text{fs}} = 1$ V.
Given this setup, the \ac{LMS} adaptive filter calibration converged after $< 2^{28}$ gradient update steps with a step size of $6 \times 10^{-7}$. 
As in \Sec{sec:simulation}, the input signal is a quadrature signal, with a frequency $f_{\mathrm{BP}} = f_n + \omega_{\mathcal{B}} / (8 \pi)$.

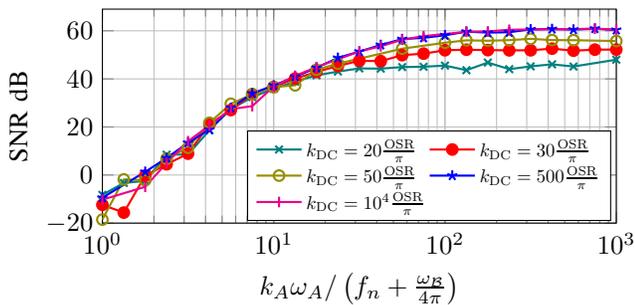
\begin{figure}
    \centering
    \input{figures/op-amp_snr_vs_GBWP}
    \caption{\label{fig:SNR_GBWP}%
    Simulated \ac{SNR}, as a function of the op-amps \ac{GBWP} and DC-gain, for a circuit implementation as in \Fig{fig:quadrature_impl},. 
    Each realization was simulated using Ngspice and the digital estimator's filter coefficients were calibrated as in \Fig{fig:adaptive_calibration}.
    }
\end{figure}

\Fig{fig:SNR_GBWP} shows the resulting \acp{SNR}, evaluated directly on the corresponding testing \ac{PSD}, as a function of \ac{GBWP} for different DC-gains. 
The DC-gains are expressed relative to $\mathrm{OSR}/\pi$ as this is the required gain-per-stage, at the bandwidth frequency, to meet the target \ac{SNR} for a given $N$ \cite{FMWLY:2023}.
The results indicate that a \ac{GBWP} less than two orders of magnitude above the signal band of interest and a DC-gain less than $500 \mathrm{OSR} / \pi$ significantly degrades the overall \ac{SNR} performance.

\begin{figure}
    \centering
    \input{figures/op-amp_psd}    
    \caption{\label{fig:PSD_GBWP}%
    \acp{PSD} demonstrating the degradation due to insufficient \ac{GBWP} and DC-gain 
    where line colors match the DC-gains from \Fig{fig:SNR_GBWP}, i.e., the teal colored \ac{PSD} correspond to $\kappa_A = 20 \mathrm{OSR}/\pi$, the yellow \ac{PSD} corresponds to $\kappa_A = 50 \mathrm{OSR}/\pi$, and the blue \ac{PSD} corresponds to $\kappa_A = 500 \mathrm{OSR}/\pi$. 
    }
\end{figure}
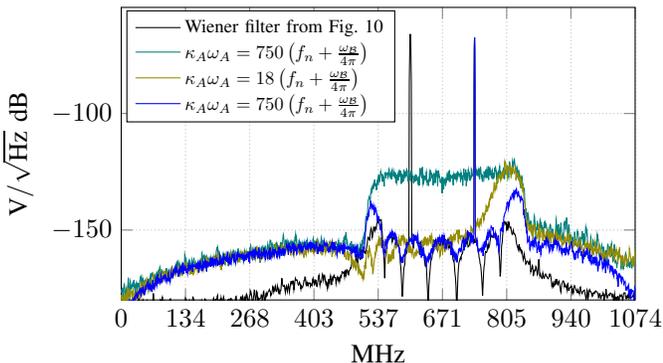

Additionally, \Fig{fig:PSD_GBWP} compares the nominal Wiener filter from \Fig{fig:psd} with a selection of the calibrated estimates from \Fig{fig:SNR_GBWP}.
\Fig{fig:PSD_GBWP} shows that an insufficient DC-gain, i.e., the teal-colored $20 \mathrm{OSR} / \pi$ \ac{PSD}, results in a significantly increased in-band noise floor. Similarly, an insufficient \ac{GBWP}, i.e., the yellow-colored \ac{PSD}, fails to suppress the conversion error towards the upper bandwidth edge. However, for a sufficient \ac{GBWP} and DC-gain, i.e., the blue-colored \ac{PSD}, the successful calibration is evident as the characteristic leapfrog ripples emerge from the in-band noise floor.
The generally rising noise floor towards the bandwidth edges, also noticeable for a sufficient \ac{GBWP} and DC-gain, i.e., the blue-colored \ac{PSD}, is due to the leapfrog parametrization in \Sec{sec:leapfrog_parametrization} that for $N < \infty$ results in a marginally smaller bandwidth than $\omega_\mathcal{B}$. 
Subsequently, as the reference filter $h_0[k]$ was chosen to have (-3dB) amplitude gain at exactly $\omega_\mathcal{B}$, this results in an increased conversion error around the bandwidth edges. 
This could be circumvented by adjusting the reference filter $h_0[k]$ alternatively increasing $\alpha$ in \Eq{eq:leapfrog_relation}.
Incidentally, this is also the reason for the lower reported \ac{SNR} in the case of calibration, see \Fig{fig:SNR_GBWP}, compared to those of the Wiener filter solution.
Additionally, a slightly lower signal peak than for the Wiener filter solution is due to the reduced input signal.
Finally, the calibrated filters amplitude responses, for $k_A \omega_A (f_n + \frac{\omega_\mathcal{B}}{4 \pi}) = 750$ and $\kappa_A = 10^4$, are shown in \Fig{fig:calibrated_filters}.

\section{Conclusions}\label{sec:conlcusions}

The \ac{QCBADC} design principle extends any low-pass \acp{CBADC} into a quadrature version centered around a desired notch frequency without loss of \ac{SNR} or stability margin. Nominal performance and tuning range are verified by behavioral simulations of different system orders and \acp{OSR}.
The introduction of digital calibration simplifies the analog design task as the analog transfer function and final state swing then determine the nominal ADC performance. 
Robustness is demonstrated by behavioral Monte Carlo simulations, with up to $10$\% component variations, resulting in a performance range of approximately $(-4, 2)$ dB, relative to the nominal \ac{SNR}. Electrical simulations confirm graceful \ac{SNR} degradation for a first-order op-amp implementation with insufficient \ac{GBWP}. These results demonstrate that \ac{QCBADC} is a good alternative for implementing RF digitizers with continuous tuning of the notch frequency, opening the doors to more efficient realization of software-defined radios.

\appendix 
\label{app:appendix}
\subsection{The Signal Estimate Decomposition}\label{app:decomposition}

\subsubsection{Scalar Control Signal Case}\label{app:decomposition_scalar_case}
The discrete-time convolution can be written as 
\begin{IEEEeqnarray}{rCl}
    \hat{u}[k] & \eqdef & \sum_{k_1 \in \Z} s[k_1] h[k - k_1] \label{eq:discrete_continuous_first} \\
          &=& \sum_{k_1 \in \Z} s[k_1] (\tilde{g} \ast g \ast \mathrm{DAC})((k - k_1) T_s)  \label{eq:discrete_continuous_second}  \\
          &=& \int (\tilde{g} \ast g)(\tau)  \sum_{k_1 \in \Z} s[k_1] \mathrm{DAC}(kT - \tau - k_1 T_s ) \dd \tau \nonumber \\ \label{eq:sum_integral_change} \\
        &=& \int (\tilde{g} \ast g)(\tau) s(kT_s - \tau) \dd \tau \label{eq:discrete_continuous_second_to_last} \\
        &=& \left(\tilde{g} \ast g \ast s\right)(kT_s) \label{eq:discrete_continuous_last}
\end{IEEEeqnarray}
where the steps to \Eq{eq:discrete_continuous_second} and \Eq{eq:discrete_continuous_second_to_last} follows from the definitions in \Eq{eq:ct_filter_coefficients} and
\Eq{eq:ct_control_signal} respectively.
Finally, by using the relation in \Eq{eq:ct_observation}, 
\Eq{eq:discrete_continuous_last} results in 
\begin{IEEEeqnarray}{rCl}
    \left(\tilde{g} \ast g \ast s\right)(kT_s) & = & \int \tilde{g}(kT_s) (g \ast s)(\tau - kT_s) \dd \tau \\ 
    & = & \int \tilde{g}(kT_s) \left(g \ast u - x\right)(kT_s - \tau) \dd \tau \nonumber \\ \\
    & = & \int \tilde{g}(kT_s) \left(g \ast u\right)(kT_s - \tau) \dd \tau \nonumber \\
    &   & - \int \tilde{g}(kT_s) x(kT_s - \tau) \dd \tau \label{eq:continuous_derivation_step2} \\
    & = & \left(\tilde{g} \ast g \ast u\right)(kT_s) - \left(\tilde{g} \ast x \right)(kT_s)
\end{IEEEeqnarray}
which gives \Eq{eq:ctsdm_decomp}.

Admittedly, several mathematical subtleties must be addressed for a general $\tilde{g}(t)$ impulse response.
In particular, the exchange of the summation and integration in \Eq{eq:sum_integral_change} assumes both $(\tilde{g} \ast g)(t)$ and $\mathrm{DAC}(t)$ to be absolutely integrable. Furthermore, we assume $\left(\tilde{g} \ast g \ast s\right)(kT)$, $\left(\tilde{g} \ast g \ast u\right)(kT)$, and $\left(\tilde{g} \ast x\right)(kT)$ to be well-defined convolutions. However, as $h[k]$, i.e. $(\tilde{g} \ast g \ast \mathrm{DAC})(k T)$, cf. \Eq{eq:ct_filter_coefficients}, is a digital filtering design choice, the resulting discrete-time convolution \Eq{eq:ct_estimate} can be assumed well defined, and thus these mathematical practicalities are, for reasonable choices of $h[k]$, of limited practical concern.

\subsubsection{Multiple Control Signals Case}\label{app:sampling_decomposition_multi_controls}
Similar to the steps in \App{app:decomposition_scalar_case}, the multi-control signal estimation can be deduced by the following steps:
\begin{IEEEeqnarray}{rCl}
    \hat{u}[k] & \eqdef & \sum_{\ell=1}^N \sum_{k_1 \in \Z} s_\ell[k_1] h_\ell[k - k_1] \\
    & = &  \sum_{\ell=1}^N \sum_{k_1 \in \Z} \kappa s_\ell[k_1] (\tilde{g} \ast g_\ell \ast \theta)((k - k_1)T_s) \label{eq:applying_multi_control_definition} \\
    & = & \sum_{\ell=1}^N \int (\tilde{g} \ast g_\ell)(\tau) s_\ell(k T_s - \tau) \dd \tau \label{eq:multi_control_temp1} \\
    & = & \sum_{\ell=1}^N (\tilde{g} \ast g_\ell \ast s_\ell)(k T_s) \label{eq:multi_control_temp2}
\end{IEEEeqnarray}
where \Eq{eq:applying_multi_control_definition} follows from substituting \Eq{eq:discrete_filter_definition} and \Eq{eq:multi_control_temp1} follows the steps in going from \Eq{eq:discrete_continuous_second} to \Eq{eq:discrete_continuous_second_to_last}. 
By substituting \Eq{eq:general_cbadc_observation} into \Eq{eq:multi_control_temp2} 
\begin{IEEEeqnarray}{rCl}
    \IEEEeqnarraymulticol{3}{l}{
        \sum_{\ell=1}^N \left(\tilde{g} \ast g_\ell \ast s_\ell\right)(kT_s) 
        }  \nonumber\\* \quad
    & = & \sum_{\ell=1}^N \int \tilde{g}(kT_s) (g_\ell \ast s_\ell)(\tau - kT_s) \dd \tau \\ 
    & = & \int \tilde{g}(kT_s) \sum_{\ell=1}^N (g_\ell \ast s_\ell)(\tau - kT_s) \dd \tau \\ 
    & = & \int \tilde{g}(kT_s) \left(g \ast u - x\right)(kT_s - \tau) \dd \tau \\
    & = & \left(\tilde{g} \ast g \ast u\right)(kT_s) - \left(\tilde{g} \ast x \right)(kT_s)
\end{IEEEeqnarray}
we end up with the sampling expression as in \Eq{eq:sampling}.

\subsection{Local Quadrature Digital-Control}\label{app:lqdc}
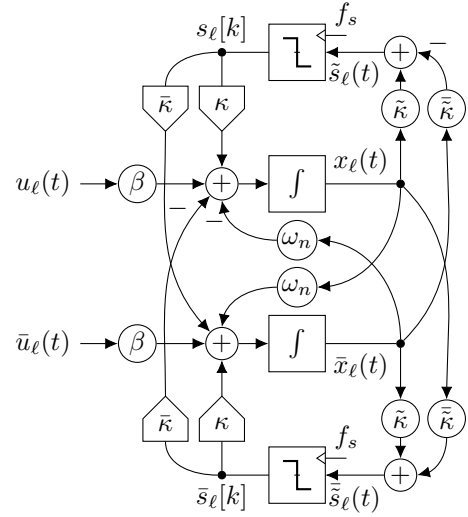
\begin{figure}
    \centering
    \input{figures/oscillator_AF}
    \caption{\label{fig:local_oscillator}%
    An analog oscillator with an quadrature input pair $u_\ell(t), \bar{u}_\ell(t)$, and a local quadrature digital control. 
    }
\end{figure}

The worst-case pair-wise quadrature state system, excluding the possibility of local positive feedback, is shown in \Fig{fig:local_oscillator}, i.e., an oscillator stabilized with local digital control.
In the low-pass analogy, an uncontrolled state's magnitude increases the most for a constant input signal during a control period $T_s$. 
Equivalently, an oscillating signal with a frequency $\omega_n$ increases the state norm the most during a control period.
This can be seen when solving the corresponding system of differential equations for the oscillator in \Fig{fig:local_oscillator}, for $s_\ell(t) = \bar{s}_\ell(t) = 0$ as
\begin{IEEEeqnarray}{rCl}
    \vct{u}_\ell(t) & \eqdef & \mat{\Theta}(\omega_n t) \begin{pmatrix}u_\ell(0) \\ \bar{u}_\ell(0) \end{pmatrix} \\
    \vct{x}_{\ell}(T_s)\big|_{s = 0} & = & \beta \int_0^{T_s} \exp\left({\begin{pmatrix}0 & -\omega_n \\ \omega_n & 0 \end{pmatrix}(T_s - \tau)} \right) \vct{u}_\ell(\tau) \dd \tau \nonumber \\ \\
    & = & \beta T_s \mat{\Theta}(\omega_n T_s) \vct{u}_\ell(0) \label{eq:input_strength}
\end{IEEEeqnarray}
where $\mat{\Theta}(t)$ is a rotation matrix as defined in \Eq{eq:rotation_matrix} and the properties in \App{app:rotation_matrices} have been used extensively. 
Similarly, a pair-wise control signal, in the absence of an input signal, i.e., $u(t)=\bar{u}(t)=0$, generates the following state trajectory
\begin{IEEEeqnarray}{rCl}
    \vct{x}_\ell(T_s)\big|_{u=0} & = & \int_0^{T_s} \exp\left({\begin{pmatrix}0 & -\omega_n \\ \omega_n & 0 \end{pmatrix}(T_s - \tau)} \right) \vct{s}_\ell(\tau) \dd \tau \nonumber \\ \\
    & = & 2\frac{ \sqrt{\kappa_\phi^2 + \bar{\kappa}_\phi^2}}{\omega_n} \sin\left(\frac{\omega_n T}{2}\right) \mat{\Theta}\left(\frac{\omega_n T}{2} + \phi_\kappa \right) \vct{s}_\ell[0] \nonumber \\ \label{eq:control_strength}
\end{IEEEeqnarray}
where $\phi_\kappa \eqdef \arctan\left(\bar{\kappa}_\phi / \kappa_\phi\right)$ and we made use of properties of rotation matrices from \App{app:rotation_matrices}.
Without loss of generality, only the control period $t\in[0, T_s)$ will be considered.

Extending the local digital control conditions for integrators \cite{M:20} to the case of oscillators,
reduces to upholding three conditions; namely: each control signal strength must match that of a worst-case input signal, the local quadrature digital control needs to be self-stable in the sense that any bounded initial state pair remains bounded throughout a control period $T_s$, and finally, the worst possible superposition of control signal and input signal must remain bounded during a control period $T_s$.
In the following subsections, these conditions will be addressed sequentially; the resulting conditions will amount to a unique local quadrature digital control parametrization.

\subsubsection{Matched Signal Strengths}\label{sec:matched_signal_strengths}
The first condition can be satisfied by equating the norms of \Eq{eq:input_strength} and \Eq{eq:control_strength} which results in the condition
\begin{IEEEeqnarray}{rCl}%
    \sqrt{\kappa_\phi^2 + \bar{\kappa}_\phi^2} & \overset{!}{=} & \frac{\beta T_s \omega_n}{2 \sin\left(\frac{\omega_n T_s}{2}\right)}. \label{eq:control_gain}
\end{IEEEeqnarray}
where we have assumed the same bound on both control and input signal, i.e., $\|\mat{\Theta}(\phi)\vct{u}_\ell(t)\|_2 = \|\vct{u}_\ell(t)\|_2 = \|\vct{s}_\ell(t)\|_2$.

\subsubsection{Self Stability}
Anticipating how a state trajectory evolves over a control period $T_s$ and aligning the control response accordingly ensures self-stability.
Specifically, an initial state will evolve as
\begin{IEEEeqnarray}{rCl}
    \vct{x}_\ell(T_s) & = & \mat{\Theta}(\omega_n T_s) \vct{x}_\ell(0). \label{eq:state_evolution}
\end{IEEEeqnarray}
From \Sec{sec:matched_signal_strengths} follows that the largest permissible initial state pair would result after a control period $T_s$ when both control signal and input signal superimpose. The largest permissible initial state pair $\vct{x}_\ell(0)$ thus follows as twice that of \Eq{eq:control_strength}
assuming the initial state was $\vct{x}_\ell(-T_s) = \begin{pmatrix} 0, 0\end{pmatrix}^\T$.

Furthermore, the binary control signals are configured such that, when faced with the largest permissible initial state pair, the quantized control observation and control signal are identical, i.e.,
$\vct{s}[0] = \tilde{s}(0)$. 
Equating \Eq{eq:state_evolution}, with the initial state as twice that of \Eq{eq:control_strength}, with the negative corresponding control contribution, i.e., \Eq{eq:control_observation} and \Eq{eq:control_strength} and rearranging results in the two conditions
\begin{IEEEeqnarray}{rCl}%
    \sqrt{\tilde{\kappa}_\phi^2 + \bar{\tilde{\kappa}}_\phi^2} & \overset{!}{=} & \frac{\omega_n}{2 \sqrt{\kappa_\phi^2 + \bar{\kappa}_\phi^2} \sin\left(\frac{\omega_n T_s}{2}\right)} \label{eq:kappa_tilde_scaling} \\
    \arctan\left(\frac{\bar{\tilde{\kappa}}_\phi}{\tilde{\kappa}_\phi}\right) & \overset{!}{=} & \omega_n \left(\frac{T_s}{2} + \tau_{\text{DC}}\right) - \phi_\kappa + \pi. \label{eq:kappa_tilde_angle}
\end{IEEEeqnarray}

\subsubsection{Worst-Case Superposition}
Similarly to the chain-of-integrators case \cite{M:20}, 
\begin{IEEEeqnarray}{rCl}%
    2 \beta T_s & \leq & 1 \label{eq:superposition}
\end{IEEEeqnarray}
will ensure that the state vector is bounded for a worst-case input and control signal superposition.

By combining \Eq{eq:control_gain} and  
\Eq{eq:kappa_tilde_scaling}-\Eq{eq:superposition} follows the parametrization in
\Eq{eq:kappa_phi}-\Eq{eq:control_rate}.

\subsection{Rotation Matrices}\label{app:rotation_matrices}
A rotation matrix is defined as
\begin{IEEEeqnarray}{rCl}%
    \mat{\Theta}(\phi) & \eqdef & \begin{pmatrix}\cos(\phi) & -\sin(\phi) \\ \sin(\phi) & \cos(\phi)\end{pmatrix}.  \labell{eq:rotation_matrix}%
\end{IEEEeqnarray}
They are commutative
\begin{IEEEeqnarray}{rCl}%
    \mat{\Theta}(\phi_1)\mat{\Theta}(\phi_2) & = & \mat{\Theta}(\phi_1 + \phi_2) \\
                                             & = & \mat{\Theta}(\phi_2)\mat{\Theta}(\phi_1). \labell{eq:rotation_matrix_commutative}%
\end{IEEEeqnarray}
There is a relation between matrix exponentials and rotation matrix (easily verified using Taylor series)
\begin{IEEEeqnarray}{rCl}%
    \mat{\Theta}(\phi) & = & \exp\left(\begin{pmatrix}0 & -\phi \\ \phi & 0 \end{pmatrix}\right). \label{eq:rotation_exponential}
\end{IEEEeqnarray}
A negation is a rotation
\begin{IEEEeqnarray}{rCl}%
    - \mat{\Theta}(\phi)
        & = & \mat{\Theta}(\phi + \pi).
\end{IEEEeqnarray}
The integration over a rotation matrix follows as 
\begin{IEEEeqnarray}{rCl}%
    \int_{-\Delta}^{\Delta} \mat{\Theta}(\phi \tau) \dd \tau 
    & = & \frac{2}{\phi} \sin\left(\phi \Delta \right) \mat{\Theta}\left(\pi \right) \label{eq:rotation_matrix_integral_2} 
\end{IEEEeqnarray}
where we have used the property
\begin{IEEEeqnarray}{rCl}%
    \mat{\Theta}(\phi) - \mat{\Theta}(-\phi)  
        & = & 2 \sin(\phi) \mat{\Theta}\left(\frac{\pi}{2}\right). \label{eq:rotation_matrix_difference} 
\end{IEEEeqnarray}
Furthermore,
\begin{IEEEeqnarray}{rCl}%
    \begin{pmatrix}a & -b \\ b & a\end{pmatrix} = \sqrt{a^2 + b^2} \, \mat{\Theta}\left(\arctan\left(\frac{b}{a}\right)\right) \label{eq:arctan_rotation_matrix}
\end{IEEEeqnarray}
for any $a,b \in \R$.

\newcommand{\norcas}{IEEE Nordic Circuits and Syst. Conf. (NorCAS)}
\newcommand{\mwscas}{IEEE Int. Midwest Symp. Circuits and Syst. (MWSCAS)}
\newcommand{\iscas}{Proc. IEEE Int. Symp. Circuits Syst. (ISCAS)}
\newcommand{\isscc}{IEEE Int. Solid-State Circuits Conf. (ISSCC)}
\newcommand{\ita}{Information Theory \& Applications Workshop (ITA)}

\newcommand{\tcasi}{IEEE Trans. Circuits Syst. I: Reg. Papers}
\newcommand{\tcasii}{IEEE Trans. Circuits Syst. II: Exp. Briefs}
\newcommand{\procIEEE}{Proceedings of the IEEE}
\newcommand{\vlsi}{IEEE Trans. Very Large Scale Integration (VLSI) Systems}

\newcommand{\jssc}{IEEE J. Solid-State Circuits}
\newcommand{\jestcs}{IEEE J. Emerg. Sel. Topics Circuits Syst.}

\newcommand{\phd}[3]{#1, \enquote{#2,} Ph.D. dissertation, #3.}

\end{document}

%% file: acronyms.tex
\begin{acronym}
    \acro{CBADC}{control-bounded analog-to-digital converter}
    \acro{CTSDM}[CT-$\Sigma\Delta$M]{continuous-time sigma-delta modulator}
    \acro{CTSD}{continuous-time sigma-delta}
    \acro{DTSDM}{discrete-time sigma-delta modulator}
    \acro{CRFB}{cascade of resonators with feedback}
    \acro{CRFF}{cascade of resonators with feedforward}

    \acro{ENOB}{effective number of bits}
    \acro{SNR}{signal-to-noise ratio}
    \acro{SNDR}{signal-to-noise-and-distortion ratio}
    \acro{SQNR}{signal-to-quantization-noise ratio}
    \acro{SC}{switched-capacitor}
    \acro{CT}{continuous-time}
    \acro{DT}{discrete-time}
    \acro{PSD}{power spectral density}

    \acro{NTF}{noise transfer function}
    \acro{STF}{signal transfer function}

    \acro{CI}{chain-of-integrators}
    \acro{LF}{leapfrog}
    \acro{DC}{digital control}
    \acro{AS}{analog system}
    \acro{DE}{digital estimator}

    \acro{RZ}{return-to-zero}
    \acro{NRZ}{non-return-to-zero}
    \acro{SCR}{switched-capacitor-resistor}
    \acro{DAC}{digital-to-analog converter}
    \acro{ADC}{analog-to-digital converter}

    \acro{FIR}{finite impulse response}
    \acro{FS}{full scale}
    \acro{OSR}{oversampling ratio}
    \acro{OTA}{operational transconductance amplifier}
    \acro{PLL}{phase-locked loop}
    \acro{VCO}{voltage-controlled oscillator}
    \acro{PM}{phase-modulated}
    \acro{FM}{frequency-modulated}
    
    \acro{GBWP}{gain-bandwidth product}
    \acro{PLL}{phase-locked loop}
    \acro{LMS}{least mean squares}
    \acro{RLS}{recursive least squares}
    \acro{AF}{analog frontend}
    \acro{SDM}[$\Sigma\Delta$M]{sigma-delta modulator}
    \acro{QSDM}[Q-$\Sigma\Delta$M]{quadrature sigma-delta modulator}
    \acro{A/D}{analog-to-digital}
    \acro{BPSDM}[BP-$\Sigma\Delta$M]{band-pass sigma-delta modulator}
    \acro{RF}{radio frequency}
    \acro{CT}{continuous-time}
    \acro{DT}{discrete-time}
    \acro{SDR}{software-define-radio}
    \acro{QCBADC}[Q-CBADC]{quadrature control-bounded analog-to-digital converter}
    \acro{BW}{bandwidth}
    \acro{VRMS}{root-mean-square voltage}
    \acro{dBFS}{decibels relative to full scale}
\end{acronym}

%% file: figures/sigma_delta_ct.tex
\begin{tikzpicture}[node distance=1.25cm,AnalogMultiplier/.style={AnalogSum, font={}, minimum size=0.9cm}]

    \node (input) {$u(t)$};
    \node[AnalogSum] (sum) at ($(input) + (1.25,0)$) {};
    \node[Analog] (lti) at ($(sum)+(1,0)$) {$G(s)$};
    \draw pic[rotate=0,yscale=1,xscale=-1] (sampler) at ($(lti) + (1.25, 0)$) {sampler}; 
    \draw pic[rotate=0, yscale=1] (qunatizer) at ($(lti) + (2, 0)$) {multiBitQuantizer};
    \node[AnalogBranch] (branch) at ($(qunatizer-out)+(0.5,0)$) {};
    \node (x) at ($(branch) + (1.25,0)$) {$s[k]$};
    \draw pic[scale=2, rotate=90] (dac) at ($(lti) + (0,-1)$) {DAC={DAC}};
    \node[below] at ($(sampler-out) + (0.125,0)$) {$x(t)$};
    \node[above] at ($(sampler-in)+(-0.5, 0.175)$) {$f_s$};

    \draw[Arrow] (input) -- (sum);
    \draw[Arrow] (sum) -- (lti);
    \draw[] (lti) -- (sampler-out);
    \draw[] (qunatizer-out) -- (branch);
    \draw[Arrow] (branch) -- (x);
    \draw[Arrow] (branch) |- (dac-in);
    \draw[Arrow] (dac-out) -| node[left] {$s(t)$} (sum);

    \node at ($(sum)+(-0.25,-0.25)$) {$-$};
\end{tikzpicture}

%% file: figures/sigma_delta_estimation_model_2.tex
\begin{tikzpicture}[node distance=1.25cm,AnalogMultiplier/.style={AnalogSum, font={}, minimum size=0.9cm}]

    \node[] (input) {$u(t)$};
    \node[Analog] (input_g) at ($(input) + (1.25,0)$) {$G(s)$};
    \draw pic[rotate=0, yscale=1, xscale=-1] (sampler_1) at ($(input_g) + (1,0)$) {sampler};
    \node[AnalogSum] (sum) at ($(sampler_1-in) + (0.25,0)$) {};
    \draw pic[rotate=0,yscale=1,xscale=1] (sampler_2) at ($(sum) + (0.5,-1)$) {sampler};
    \node[Analog] (g) at ($(sampler_2-out) + (0.5,0)$) {$G(s)$};
    \draw pic[scale=2, rotate=90] (dac) at ($(g) + (1.25,0)$) {DAC={DAC}};

    \node[AnalogBranch] (branch) at ($(sum) + (3.5,0)$) {};

    \node[] (s) at ($(branch) + (1,0)$) {$s[k]$};

    \draw pic[rotate=0, yscale=1] (qunatizer) at ($(sum) + (2.25, 0)$) {multiBitQuantizer};

    \node[above] at ($(sampler_1-in)+(-0.5, 0.175)$) {$f_s$};
    \node[above] at ($(sampler_2-out)+(-0.5, 0.175)$) {$f_s$};

    \draw[Arrow] (sampler_2-in) -- (sum);
    \draw[] (g) -- (sampler_2-out);
    \draw[Arrow] (branch) |- (dac-in);
    \draw[Arrow] (dac-out) -- (g);

    \draw[Arrow] (input) -- (input_g);
    \draw[] (input_g) -- (sampler_1-out);
    \draw[Arrow] (sampler_1-in) -- (sum);
    \draw (qunatizer-out) -- (branch);
    \draw (sum) -- node[above] {$x(kT_s)$} (qunatizer-in);
    \draw[Arrow] (branch) -- (s);
\end{tikzpicture}

%% file: figures/sigma_delta_permutation_1.tex
\begin{tikzpicture}[node distance=1.25cm,AnalogMultiplier/.style={AnalogSum, font={}, minimum size=0.9cm}]

    \node[] (input) {$u(t)$};
    \node[Analog] (input_g) at ($(input) + (1.25,0)$) {$G(s)$};
    \node[AnalogSum] (sum) at ($(input_g) + (1.25,0)$) {};
    \node[Analog] (g) at ($(sum) + (0.675,-1.8125)$) {$G(s)$};
    \draw pic[scale=2, rotate=90] (dac) at ($(g) + (1.675,0)$) {DAC={DAC}};

    \node[AnalogBranch] (branch) at ($(sum) + (3.325,0)$) {};

    \draw pic[rotate=-90,yscale=1,xscale=-1] (sampler) at ($(branch) + (0, -0.5625)$) {sampler}; 
    \draw pic[rotate=-90, yscale=1] (qunatizer) at ($(sampler-in) + (0, -0.25)$) {multiBitQuantizer};

    \node[AnalogBranch] (branch_2) at ($(qunatizer-out) + (0,-0.25)$) {};
    \node[above] at ($(sampler-in)+(0.5, 0.175)$) {$f_s$};

    \node[] (x) at ($(branch) + (1,0)$) {$x(t)$};
    \node (s) at ($(branch_2) + (1,0)$) {$s[k]$};
    \draw[] (sum) -- (branch);
    \draw[Arrow] (dac-out) -- node[above] {$s(t)$} (g);
    \draw (qunatizer-out) -- (branch_2);
    \draw[Arrow] (branch_2) -- (dac-in);
    \draw[Arrow] (g) -| (sum);
    \draw[Arrow] (input_g) -- (sum);

    \draw[Arrow] (input) -- (input_g);
    \draw[Arrow] (branch) -- (x);
    \draw[Arrow] (branch_2) -- (s);

\end{tikzpicture}

%% file: figures/sigma_delta_discrete_time.tex
\begin{tikzpicture}[node distance=1.25cm,AnalogMultiplier/.style={AnalogSum, font={}, minimum size=0.9cm}]

    \node (input) {$U(i \omega)$};
    \node[Analog] (input_g) at ($(input) + (1.675,0)$) {$G(i\omega)$};
    \node[AnalogSum] (sum) at ($(input_g) + (2,0)$) {};
    \node[Analog] (lti) at ($(sum)+(1.3125,0)$) {$\frac{1}{1+\bar{G}(e^{i\Omega})}$};
    \node[] (branch) at ($(lti) + (1.3125,0)$) {};
    \node[Analog] at ($(branch) + (1.3125,0)$) (aa) {$H(e^{i\Omega})$};
    \draw pic[rotate=0, yscale=1, xscale=-1] (sampler) at ($(aa) + (1.125, 0)$) {sampler};
    \node[above,inner sep=0] (down-sampling) at ($(sampler-in)+(-0.275, 0.175)$) {$f_s/\mathrm{OSR}$};
    \node[draw=gray,inner sep=2pt,thick,dashed,fit=(aa)(sampler-in)(down-sampling)] (decimation) {};
    \node (est) at ($(sampler-in) + (1.3125,0)$) {$\hat{U}(e^{i\Omega})$};

    \node[below] at ($(decimation) + (0,-0.6125)$) {Decimation filter};

    \node[above] at (branch) {$S(e^{i\Omega})$};
    \node[above] at ($(input_g) + (1.25,0)$) {$U_G(e^{i\Omega})$};

    \node (q) at ($(sum)+(0,1)$) {$Q(e^{i\Omega})$};
    \draw[Arrow] (q) -- (sum);

    \draw[Arrow] (input) -- (input_g);
    \draw[Arrow] (input_g) -- (sum);
    \draw[Arrow] (sum) -- (lti);
    \draw[Arrow] (lti) -- (aa);
    \draw (aa) -- (sampler-out);
    \draw[Arrow] (sampler-in) -- (est);

\end{tikzpicture}

%% file: figures/cbadc_estimation.tex
\begin{tikzpicture}[node distance=1.25cm,AnalogMultiplier/.style={AnalogSum, font={}, minimum size=0.9cm}]

    \node (input) {$U(i\omega)$};
    \node[Analog] (input_g) at ($(input) + (1.5,0)$) {$G(i\omega)$};
    \node[AnalogSum] (sum) at ($(input_g) + (1.25,0)$) {};
    \node[Analog] (lti) at ($(sum)+(2.75,0)$) {$\tilde{G}(i\omega)$};
    \node (quantization_noise) at ($(sum) + (0,1)$) {$-X(i\omega)$};
    \draw pic[rotate=0, yscale=1, xscale=-1] (sampler) at ($(lti) + (1.125, 0)$) {sampler};
    \node[above,inner sep=0] (down-sampling) at ($(sampler-in)+(-0.275, 0.175)$) {$f_s/\mathrm{OSR}$};
    \node (est) at ($(sampler-in) + (1.375,0)$) {$\hat{U}(e^{i\Omega})$};
    \node[draw=gray,inner sep=2pt,thick,dashed,fit=(lti)(sampler-in)(down-sampling)] (decimation) {};

    \node[below] at ($(decimation) + (0,-0.6125)$) {Digital Estimator};

    \node[above] at ($(sum) + (1.125,0)$) {$G(i\omega) S(i\omega)$};

    \draw[Arrow] (input) -- (input_g);
    \draw[Arrow] (input_g) -- (sum);
    \draw[Arrow] (quantization_noise) -- (sum);
    \draw[Arrow] (sum) --  (lti);
    \draw (lti) -- (sampler-out);
    \draw[Arrow] (sampler-in) -- (est);
\end{tikzpicture}

%% file: figures/leapfrog_baseband.tex
\begin{tikzpicture}[node distance=1.25cm]

    \node (input_I) {$u(t)$};
    \node[AnalogMultiplier, right of=input_I, node distance=1.1cm] (beta_I_1) {$\beta$} edge[BackArrow] (input_I);
    \node[AnalogSum, right of=beta_I_1, node distance=0.875cm] (sum_I_1) {} edge[BackArrow] (beta_I_1);
    \node[Analog, right of=sum_I_1, node distance=1cm] (int_I_1) {$\int$} edge[BackArrow] (sum_I_1);
    \node[AnalogBranch, right of=int_I_1, node distance=1.375cm] (x_I_1) {} edge (int_I_1);
    \node[anchor=south] at (x_I_1) {$x_1(t)$};

    \node[AnalogMultiplier, right of=x_I_1, node distance=1.25cm] (beta2) {$\beta$} edge[BackArrow] (x_I_1);
    \node[AnalogSum, right of=beta2, node distance=0.875cm] (sum_I_2) {} edge[BackArrow] (beta2);
    \node[Analog, right of=sum_I_2, node distance=1cm] (int_I_2) {$\int$} edge[BackArrow] (sum_I_2);
    \node[AnalogBranch, right of=int_I_2, node distance=1.375cm] (x_I_2) {} edge (int_I_2);
    \node[anchor=south west] at (x_I_2) {$x_2(t)$};
    
    \node[AnalogMultiplier] (alpha_I_1) at ($ (beta2) + (-0,1.1)$) {$\alpha$};
    \draw[Arrow] (x_I_2) to[out=135, in=0] (alpha_I_1);
    \draw[Arrow] (alpha_I_1) to[out=180, in=45] (sum_I_1);

    \node[AnalogMultiplier, right of=x_I_2, node distance=1.25cm] (beta3) {$\beta$} edge[BackArrow] (x_I_2);
    \node[AnalogSum, right of=beta3, node distance=0.875cm] (sum_I_3) {} edge[BackArrow] (beta3);
    \node[Analog, right of=sum_I_3, node distance=1cm] (int_I_3) {$\int$} edge[BackArrow] (sum_I_3);
    \node[AnalogBranch, right of=int_I_3, node distance=1.375cm] (x_I_3) {} edge (int_I_3);
    \node[anchor=south west] at (x_I_3) {$x_3(t)$};

    \node[AnalogMultiplier] (alpha_I_2) at ($ (beta3) + (-0,1.1)$) {$\alpha$};
    \draw[Arrow] (x_I_3) to[out=135, in=0] (alpha_I_2);
    \draw[Arrow] (alpha_I_2) to[out=180, in=45] (sum_I_2);

    \node[right of=x_I_3, node distance=0.8275cm] (cont_dots) {$\cdots$};
    \draw (x_I_3) -- ++(0.25,0);
    \node[AnalogMultiplier, right of=x_I_3, node distance=2cm] (betaN) {$\beta$} edge[BackArrow] (cont_dots);
    \node[AnalogSum, right of=betaN, node distance=0.875cm] (sum_I_N) {} edge[BackArrow] (betaN);
    \node[Analog, right of=sum_I_N, node distance=1cm] (int_I_N) {$\int$} edge[BackArrow] (sum_I_N);
    \node[AnalogBranch, right of=int_I_N, node distance=1cm] (x_I_N) {} edge (int_I_N);
    \node[anchor=south west] (x_i_N_label) at (x_I_N) {$x_N(t)$};
    
    \node[AnalogMultiplier] (alpha_I_N) at ($ (betaN) + (-0,1.1)$) {$\alpha$};
    \draw[Arrow] (x_I_N) to[out=135, in=0] (alpha_I_N);
    \draw (alpha_I_N) to[out=180, in=0] ++(-0.75,0) ++(-0.5,0) node[anchor=east] {$\cdots$};
    \draw[BackArrow] (sum_I_3) to[out=45, in=180] ++(2,1.1);

    \draw pic[rotate=180, yscale=-1] (quantizer1) at ($(int_I_1) + (0, -1.75)$) {quantizer};
    \draw pic[scale=1.2] (kappa1) at ($(sum_I_1) + (0,-1)$) {DAC={$\kappa$}};
    \draw[Arrow] (kappa1-out) -- node[left] {$s_1(t)$} (sum_I_1);
    \draw[Arrow] (quantizer1-out) -| (kappa1-in);
    \draw[Arrow] (x_I_1) |- (quantizer1-in);

    \draw pic[rotate=180, yscale=-1] (quantizer2) at ($(int_I_2) + (0, -1.75)$) {quantizer};
    \draw pic[scale=1.2] (kappa2) at ($(sum_I_2) + (0,-1)$) {DAC={$\kappa$}};
    \draw[Arrow] (kappa2-out) -- node[left] {$s_2(t)$} (sum_I_2);
    \draw[Arrow] (quantizer2-out) -| (kappa2-in);
    \draw[Arrow] (x_I_2) |- (quantizer2-in);

    \draw pic[rotate=180, yscale=-1] (quantizer3) at ($(int_I_3) + (0, -1.75)$) {quantizer};
    \draw pic[scale=1.2] (kappa3) at ($(sum_I_3) + (0,-1)$) {DAC={$\kappa$}};
    \draw[Arrow] (kappa3-out) -- node[left] {$s_3(t)$} (sum_I_3);
    \draw[Arrow] (quantizer3-out) -| (kappa3-in);
    \draw[Arrow] (x_I_3) |- (quantizer3-in);

    \draw pic[rotate=180, yscale=-1] (quantizerN) at ($(int_I_N) + (0, -1.75)$) {quantizer};
    \draw pic[scale=1.2] (kappaN) at ($(sum_I_N) + (0,-1)$) {DAC={$\kappa$}};
    \draw[Arrow] (kappaN-out) -- node[left] {$s_N(t)$} (sum_I_N);
    \draw[Arrow] (quantizerN-out) -| (kappaN-in);
    \draw[Arrow] (x_I_N) |- (quantizerN-in);
    
\end{tikzpicture}

%% file: figures/adaptive_calibration.tex
\begin{tikzpicture}[node distance=1.25cm]
    \node[Analog] (analog_frontend) {AF};
    \node[Analog] (adaptive_filter) at ($(analog_frontend) + (4.125,0)$) {$\sum_{\ell = 1}^N  (h_\ell \star s_\ell)[k]$};
    \node[AnalogSum] (asum) at ($(adaptive_filter) + (2, 0)$) {};
    \coordinate (abranch) at ($(asum) + (0.5, 0)$) {};
    \coordinate (dac) at ($(analog_frontend) + (-1.75,0)$) {};
    \draw pic[scale=2, rotate=-90] (dac) at (dac) {DAC={DAC}};

    \node[Analog] (desired_filter) at ($(dac) + (0.5, -1.5)$) {$(h_0 \star s_0)[k]$};
    \node[AnalogBranch] (branch2) at ($(dac) + (-1,0)$) {};
    \node (s_0) at ($(branch2) + (-0.75, 0)$) {$s_0[k]$};

    \draw[Arrow] (analog_frontend) -- node[above] {$s_1[k], \dots, s_N[k]$} (adaptive_filter);
    \draw[] (s_0) -- (branch2);
    \draw[Arrow] (branch2) -- (dac-in);
    \draw[Arrow] (dac-out) -- node[above] {$s_0(t)$} (analog_frontend);
    \draw[Arrow] (branch2) |- (desired_filter);
    \draw (abranch) |- (asum);
    \draw[BackArrow] (asum) -- ++(-0.5, 0) -- (adaptive_filter);
    \draw[Arrow] (desired_filter) -| (asum);

    \draw[] (abranch) |- ($(adaptive_filter) + (-0.75, -1)$) -- ++(0.5, 0.625);
    \draw[Arrow] ($(adaptive_filter) + (0.25, 0.375)$) -- ++ (0.5, 0.625);

\end{tikzpicture}

%% file: figures/quadrature_extension.tex
\begin{tikzpicture}[node distance=1.25cm]
    \node[] (r_input) {$u(t)$};
    \node[Analog] (r_B) at ($(r_input) + (1.25,0)$) {$\mat{B}_{\text{LP}}$};
    \node[AnalogSum] (r_sum) at ($(r_B)+(1,0)$) {};
    \node[] (r_s) at ($(r_B) + (0,1)$) {$\vct{s}(t)$};
    \node[Analog] (r_int) at ($(r_sum) + (1,0)$) {$\int$};
    \node[Analog] (r_A) at ($(r_int) + (0,1)$) {$\mat{A}_{\text{LP}}$};
    \node[AnalogBranch] (r_branch) at ($(r_int) + (1.375,0)$) {};
    \node[right] (r_x) at ($(r_branch) + (0,0)$) {$\vct{x}(t)$};

    \node[] (i_input) at ($(r_input) + (0,-3)$) {$\bar{u}(t)$};
    \node[Analog] (i_B) at ($(i_input) + (1.25,0)$) {$\mat{B}_{\text{LP}}$};
    \node[AnalogSum] (i_sum) at ($(i_B)+(1,0)$) {};
    \node[] (i_s) at ($(i_B) + (0,-1)$) {$\bar{\vct{s}}(t)$};
    \node[Analog] (i_int) at ($(i_sum) + (1,0)$) {$\int$};
    \node[Analog] (i_A) at ($(i_int) + (0,-1)$) {$\mat{A}_{\text{LP}}$};
    \node[AnalogBranch] (i_branch) at ($(i_int) + (1.375,0)$) {};
    \node[right] (i_x) at ($(i_branch) + (0,0)$) {$\bar{\vct{x}}(t)$};

    \node[Analog] (r_wp) at ($(r_int) + (0,-1)$) {$\omega_n \mat{I}_N$};
    \node[Analog] (i_wp) at ($(i_int) + (0,1)$) {$\omega_n \mat{I}_N$};
    \node[] at ($(r_sum) + (-0.25, -0.325)$) {$-$};

    \draw[Arrow] (r_input) -- (r_B);
    \draw[Arrow] (r_B) -- (r_sum);
    \draw[Arrow] (r_wp) to[out=180, in=-90] (r_sum);
    \draw[Arrow] (r_A) to[out=180, in=90] (r_sum);
    \draw[Arrow] (r_s) -- (r_sum);
    \draw[Arrow] (r_sum) -- (r_int);
    \draw[] (r_int) -- (r_branch);
    \draw[Arrow] (r_branch) to[out=90, in=0] (r_A);
    \draw[Arrow] (r_branch) to[out=-90, in=0] (i_wp);

    \draw[Arrow] (i_input) -- (i_B);
    \draw[Arrow] (i_B) -- (i_sum);
    \draw[Arrow] (i_wp) to[out=180, in=90] (i_sum);
    \draw[Arrow] (i_A) to[out=180, in=-90] (i_sum);
    \draw[Arrow] (i_s) -- (i_sum);
    \draw[Arrow] (i_sum) -- (i_int);
    \draw[] (i_int) -- (i_branch);
    \draw[Arrow] (i_branch) to[out=-90, in=0] (i_A);
    \draw[Arrow] (i_branch) to[out=90, in=0] (r_wp);

\end{tikzpicture}

%% file: figures/quadrature_dc.tex
\begin{tikzpicture}[node distance=1.25cm,
]

    \node (x_tilde_I) {$x_\ell(.)$};
    \node (x_tilde_Q) at ($(x_tilde_I) + (0, -1)$) {$\bar{x}_\ell(.)$};

    \node[AnalogBranch] (b1) at ($(x_tilde_I) + (1,0)$) {} edge[] (x_tilde_I);
    \node[AnalogBranch] (b2) at ($(x_tilde_Q) + (1,0)$) {} edge[] (x_tilde_Q);
    
    \node[AnalogMultiplier] (m1) at ($(b1) + (1.875,0.75)$) {$\tilde{\kappa}_\phi$};
    \node[AnalogMultiplier] (m2) at ($(b1) + (1.875,0)$) {$\bar{\tilde{\kappa}}_\phi$};
    \node[AnalogMultiplier] (m3) at ($(b1) + (1.875,-1)$) {$\tilde{\kappa}_\phi$};
    \node[AnalogMultiplier] (m4) at ($(b1) + (1.875,-1.75)$) {$\bar{\tilde{\kappa}}_\phi$};

    \node[AnalogSum] (sum1) at ($(m2) + (1.5, 0)$) {};
    \node[AnalogSum] (sum2) at ($(m3) + (1.5, 0)$) {};
    \node[] at ($(sum1) + (-0.375,-0.15)$) {$-$};

    \draw[Arrow] (m1) to[out=0,in=135] (sum1);
    \draw[Arrow] (m2) -- (sum1);
    \draw[Arrow] (m3) -- (sum2);
    \draw[Arrow] (m4) to[out=0,in=-135] (sum2);

    \draw[Arrow] (b1) to[out=45,in=180] (m1);
    \draw[Arrow] (b1) to[out=-45,in=180] (m4);
    \draw[Arrow] (b2) to[out=45,in=180] (m2);
    \draw[Arrow] (b2) -- (m3);
    
    \draw pic[rotate=0, yscale=1] (quantizer1) at ($(sum1) + (1.75, 0)$) {quantizer};
    
    \draw[Arrow] (sum1) -- (quantizer1-in);
    
    \draw pic[rotate=0, yscale=1] (quantizer2) at ($(sum2) + (1.75, 0)$) {quantizer};
    
    \draw[Arrow] (sum2) -- (quantizer2-in);

    \node[AnalogBranch] (b6) at ($(quantizer1) + (1.25,0)$) {};
    \node[AnalogBranch] (b7) at ($(quantizer2) + (1.25,0)$) {};
    \draw[] (quantizer1-out) -- node[below] {$s_\ell[.]$} (b6);
    \draw[] (quantizer2-out) -- node[below] {$\bar{s}_\ell[.]$} (b7);

    \draw pic[scale=1.5,rotate=-90] (m5) at ($(b6) + (1.875,1)$) {DAC={$\kappa_\phi$}};
    \draw pic[scale=1.5,rotate=-90] (m6) at ($(b6) + (1.875,0)$) {DAC={$\bar{\kappa}_\phi$}};
    \draw pic[scale=1.5,rotate=-90] (m7) at ($(b7) + (1.875,0)$) {DAC={$\kappa_\phi$}};
    \draw pic[scale=1.5,rotate=-90] (m8) at ($(b7) + (1.875,-1)$) {DAC={$\bar{\kappa}_\phi$}};

    \node[AnalogSum] (sum3) at ($(m6-out) + (1, 0)$) {};
    \node[AnalogSum] (sum4) at ($(m7-out) + (1, 0)$) {};

    \node[] at ($(sum3) + (-0.375,-0.15)$) {$-$};

    \draw[Arrow] (m5-out) to[out=0,in=90] (sum3);
    \draw[Arrow] (m6-out) -- (sum3);
    \draw[Arrow] (m7-out) -- (sum4);
    \draw[Arrow] (m8-out) to[out=0,in=-90] (sum4);

    \draw[Arrow] (b6) to[out=45,in=180] (m5-in);
    \draw[Arrow] (b6) to[out=-45,in=180] (m8-in);
    \draw[Arrow] (b7) to[out=45,in=180] (m6-in);
    \draw[Arrow] (b7) -- (m7-in);

    \node (s_I) at ($(sum3) + (1.5,0)$) {$s_\ell(.)$} edge[BackArrow] (sum3);
    \node (s_Q) at ($(sum4) + (1.5,0)$) {$\bar{s}_\ell(.)$} edge[BackArrow] (sum4);

    \node[anchor=north] at ($(sum1) + (0.615, 0)$) {$\tilde{s}_\ell(.)$};
    \node[anchor=north] at ($(sum2) + (0.615, 0)$) {$\tilde{\bar{s}}_\ell(.)$};

\end{tikzpicture}

%% file: figures/kappa_figure.tex
\begin{tikzpicture}
        \pgfplotsset{
            grid=both, 
        }
        \pgfplotsset{ylabel near ticks, xlabel near ticks}
        \pgfplotsset{ 
            legend cell align={left},
            legend style={
                at={(0.025,0.65)},
                anchor=north west,
                fill opacity=1,
                draw opacity=1,
                text opacity=1,
                nodes={scale=0.8, transform shape}
            }
        }
        \begin{axis}[
            xlabel=$\omega_p T_s / (2 \pi)$,
            ylabel=Coeff. value,
            xmin=0,
            xmax=0.5,
            ymax=1,
            width=0.95\columnwidth,
            height=\columnwidth/2,
            xtick = {0, 0.0625, 0.125, 0.1875, 0.25, 0.3125, 0.375, 0.4375, 0.5},
            xticklabels = {$0$, ,$1/8$, ,$1/4$, ,$3/8$, , $1/2$},
            ]
            \addplot[color=black, very thick] table[col sep=comma, x=fpT, y=kappa] {./figures/kappa_scaling_v2.csv};
            \addplot[color=red, very thick, densely dashed] table[col sep=comma, x=fpT, y=bar_kappa] {./figures/kappa_scaling_v2.csv};
            \addplot[color=blue, very thick, densely dotted] table[col sep=comma, x=fpT, y=tilde_kappa] {./figures/kappa_scaling_v2.csv};
            \addplot[color=olive, very thick, dash dot] table[col sep=comma, x=fpT, y=bar_tilde_kappa] {./figures/kappa_scaling_v2.csv};

            \addlegendentry{$\kappa_\phi$}
            \addlegendentry{$\bar{\kappa}_\phi$}
            \addlegendentry{$\tilde{\kappa}_\phi$}
            \addlegendentry{$\bar{\tilde{\kappa}}_\phi$}
        \end{axis}
\end{tikzpicture}

%% file: figures/quadrature_full.tex
\begin{tikzpicture}[node distance=1.25cm]

    \node (input_I) {$u(t)$};
    \node[AnalogMultiplier, right of=input_I, node distance=1.1cm] (beta_I_1) {$\beta$} edge[BackArrow] (input_I);
    \node[AnalogSum, right of=beta_I_1, node distance=0.875cm] (sum_I_1) {} edge[BackArrow] (beta_I_1);
    \node[Analog, right of=sum_I_1, node distance=1cm] (int_I_1) {$\int$} edge[BackArrow] (sum_I_1);
        \node[AnalogBranch, right of=int_I_1, node distance=1.375cm] (x_I_1) {} edge (int_I_1);

    \node[AnalogMultiplier, right of=x_I_1, node distance=1.25cm] (beta2) {$\beta$} edge[BackArrow] (x_I_1);
    \node[AnalogSum, right of=beta2, node distance=0.875cm] (sum_I_2) {} edge[BackArrow] (beta2);
    \node[Analog, right of=sum_I_2, node distance=1cm] (int_I_2) {$\int$} edge[BackArrow] (sum_I_2);
    \node[AnalogBranch, right of=int_I_2, node distance=1.375cm] (x_I_2) {} edge (int_I_2);
    \node[AnalogMultiplier] (alpha_I_1) at ($ (beta2) + (-0,1.1)$) {$\alpha$};
    \draw[Arrow] (x_I_2) to[out=135, in=0] (alpha_I_1);
    \draw[Arrow] (alpha_I_1) to[out=180, in=45] (sum_I_1);

    \node[AnalogMultiplier, right of=x_I_2, node distance=1.25cm] (beta3) {$\beta$} edge[BackArrow] (x_I_2);
    \node[AnalogSum, right of=beta3, node distance=0.875cm] (sum_I_3) {} edge[BackArrow] (beta3);
    \node[Analog, right of=sum_I_3, node distance=1cm] (int_I_3) {$\int$} edge[BackArrow] (sum_I_3);
    \node[AnalogBranch, right of=int_I_3, node distance=1.375cm] (x_I_3) {} edge (int_I_3);

    \node[AnalogMultiplier] (alpha_I_2) at ($ (beta3) + (-0,1.1)$) {$\alpha$};
    \draw[Arrow] (x_I_3) to[out=135, in=0] (alpha_I_2);
    \draw[Arrow] (alpha_I_2) to[out=180, in=45] (sum_I_2);

    \node[right of=x_I_3, node distance=0.8275cm] (cont_dots) {$\cdots$};
    \draw (x_I_3) -- ++(0.25,0);
    \node[AnalogMultiplier, right of=x_I_3, node distance=2cm] (betaN) {$\beta$} edge[BackArrow] (cont_dots);
    \node[AnalogSum, right of=betaN, node distance=0.875cm] (sum_I_N) {} edge[BackArrow] (betaN);
    \node[Analog, right of=sum_I_N, node distance=1cm] (int_I_N) {$\int$} edge[BackArrow] (sum_I_N);
    \node[AnalogBranch, right of=int_I_N, node distance=1cm] (x_I_N) {} edge (int_I_N);
    \node[anchor=south west] (x_i_N_label) at (x_I_N) {$x_N(t)$};
    
    \node[AnalogMultiplier] (alpha_I_N) at ($ (betaN) + (-0,1.1)$) {$\alpha$};
    \draw[Arrow] (x_I_N) to[out=135, in=0] (alpha_I_N);
    \draw (alpha_I_N) to[out=180, in=0] ++(-0.75,0) ++(-0.5,0) node[anchor=east] {$\cdots$};
    \draw[BackArrow] (sum_I_3) to[out=45, in=180] ++(2,1.1);

        \node (input_Q) at ($(input_I) + (0,-2.125)$)  {$\bar{u}(t)$};
        \node[AnalogMultiplier, right of=input_Q, node distance=1.1cm] (beta_Q_1) {$\beta$} edge[BackArrow] (input_Q);
        \node[AnalogSum, right of=beta_Q_1, node distance=0.875cm] (sum_Q_1) {} edge[BackArrow] (beta_Q_1);
        \node[Analog, right of=sum_Q_1, node distance=1cm] (int_Q_1) {$\int$} edge[BackArrow] (sum_Q_1);
        \node[AnalogBranch, right of=int_Q_1, node distance=1.375cm] (x_Q_1) {} edge (int_Q_1);

        \node[AnalogMultiplier, right of=x_Q_1, node distance=1.25cm] (beta2) {$\beta$} edge[BackArrow] (x_Q_1);
        \node[AnalogSum, right of=beta2, node distance=0.875cm] (sum_Q_2) {} edge[BackArrow] (beta2);
        \node[Analog, right of=sum_Q_2, node distance=1cm] (int_Q_2) {$\int$} edge[BackArrow] (sum_Q_2);
        \node[AnalogBranch, right of=int_Q_2, node distance=1.375cm] (x_Q_2) {} edge (int_Q_2);
        \node[AnalogMultiplier] (alpha_Q_1) at ($ (beta2) + (-0,-1.1)$) {$\alpha$};
        \draw[Arrow] (x_Q_2) to[out=-135, in=0] (alpha_Q_1);
        \draw[Arrow] (alpha_Q_1) to[out=180, in=-45] (sum_Q_1);

        \node[AnalogMultiplier, right of=x_Q_2, node distance=1.25cm] (beta3) {$\beta$} edge[BackArrow] (x_Q_2);
        \node[AnalogSum, right of=beta3, node distance=0.875cm] (sum_Q_3) {} edge[BackArrow] (beta3);
        \node[Analog, right of=sum_Q_3, node distance=1cm] (int_Q_3) {$\int$} edge[BackArrow] (sum_Q_3);
        \node[AnalogBranch, right of=int_Q_3, node distance=1.375cm] (x_Q_3) {} edge (int_Q_3);

        \node[AnalogMultiplier] (alpha_Q_2) at ($ (beta3) + (-0,-1.1)$) {$\alpha$};
        \draw[Arrow] (x_Q_3) to[out=-135, in=0] (alpha_Q_2);
        \draw[Arrow] (alpha_Q_2) to[out=180, in=-45] (sum_Q_2);

        \node[right of=x_Q_3, node distance=0.8275cm] (cont_dots) {$\cdots$};
        \draw (x_Q_3) -- ++(0.25,0);
        \node[AnalogMultiplier, right of=x_Q_3, node distance=2cm] (betaN) {$\beta$} edge[BackArrow] (cont_dots);
        \node[AnalogSum, right of=betaN, node distance=0.875cm] (sum_Q_N) {} edge[BackArrow] (betaN);
        \node[Analog, right of=sum_Q_N, node distance=1cm] (int_Q_N) {$\int$} edge[BackArrow] (sum_Q_N);
        \node[AnalogBranch, right of=int_Q_N, node distance=1.cm] (x_Q_N) {} edge (int_Q_N);
        \node[anchor=north west] (x_Q_N_label) at (x_Q_N) {$\bar{x}_N(t)$};

        \node[AnalogMultiplier] (alpha_Q_N) at ($ (betaN) + (-0,-1.1)$) {$\alpha$};
        \draw[Arrow] (x_Q_N) to[out=-135, in=0] (alpha_Q_N);
        \draw (alpha_Q_N) to[out=180, in=0] ++(-0.75,0) ++(-0.5,0) node[anchor=east] {$\cdots$};
        \draw[BackArrow] (sum_Q_3) to[out=-45, in=180] ++(2,-1.1);

        \node[AnalogMultiplier] (omega_1I) at ($(sum_Q_1) + (1.0 , 0.75)$) {$\omega_n$};
        \node[AnalogMultiplier] (omega_1Q) at ($(sum_Q_1) + (1.0 , 1.375)$) {$\omega_n$};

        \draw[Arrow] (x_Q_1) to[in=0, out=90] (omega_1I);
        \draw[Arrow] (omega_1I) to[out=170, in=-90] (sum_I_1);    

        \draw[Arrow] (x_I_1) to[in=0, out=-90] (omega_1Q);
        \draw[Arrow] (omega_1Q) to[out=-170, in=90] (sum_Q_1);
        \node at ($(sum_I_1) + (0.25,-0.2875)$) {$-$};

        \node[AnalogMultiplier] (omega_2I) at ($(sum_Q_2) + (1.0 , 0.75)$) {$\omega_n$};
        \node[AnalogMultiplier] (omega_2Q) at ($(sum_Q_2) + (1.0 , 1.375)$) {$\omega_n$};

        \draw[Arrow] (x_Q_2) to[in=0, out=90] (omega_2I);
        \draw[Arrow] (omega_2I) to[out=170, in=-90] (sum_I_2);    

        \draw[Arrow] (x_I_2) to[in=0, out=-90] (omega_2Q);
        \draw[Arrow] (omega_2Q) to[out=-170, in=90] (sum_Q_2);  
        \node at ($(sum_I_2) + (0.25,-0.2875)$) {$-$};

        \node[AnalogMultiplier] (omega_3I) at ($(sum_Q_3) + (1.0 , 0.75)$) {$\omega_n$};
        \node[AnalogMultiplier] (omega_3Q) at ($(sum_Q_3) + (1.0 , 1.375)$) {$\omega_n$};

        \draw[Arrow] (x_Q_3) to[in=0, out=90] (omega_3I);
        \draw[Arrow] (omega_3I) to[out=170, in=-90] (sum_I_3);    

        \draw[Arrow] (x_I_3) to[in=0, out=-90] (omega_3Q);
        \draw[Arrow] (omega_3Q) to[out=-170, in=90] (sum_Q_3);
        \node at ($(sum_I_3) + (0.25,-0.2875)$) {$-$};

        \node[AnalogMultiplier] (omega_NI) at ($(sum_Q_N) + (1.0 , 0.75)$) {$\omega_n$};
        \node[AnalogMultiplier] (omega_NQ) at ($(sum_Q_N) + (1.0 , 1.375)$) {$\omega_n$};

        \draw[Arrow] (x_Q_N) to[in=0, out=90] (omega_NI);
        \draw[Arrow] (omega_NI) to[out=170, in=-90] (sum_I_N);

        \draw[Arrow] (x_I_N) to[in=0, out=-90] (omega_NQ);
        \draw[Arrow] (omega_NQ) to[out=-170, in=90] (sum_Q_N);
        \node at ($(sum_I_N) + (0.25,-0.2875)$) {$-$};

        \draw pic[scale=1.5,rotate=180] (kappa_I_1) at ($(sum_I_1) + (0,1.6125)$) {DAC={$\kappa_\phi$}};
        \node[AnalogMultiplier] (kappa_tilde_I_1) at ($(x_I_1) + (0,1.75)$) {$\tilde{\kappa}_\phi$};
        \node[AnalogMultiplier] (kappa_tilde_bar_I_1) at ($(x_I_1) + (0.7,1.75)$) {$\bar{\tilde{\kappa}}_\phi$};
        \node[AnalogSum] (s_tilde_I_sum_1) at ($(kappa_tilde_I_1) + (0,1)$) {};
        \draw pic[rotate=0, yscale=1, xscale=-1] (q_S_I_1) at ($(s_tilde_I_sum_1) + (-1.375, 0)$) {quantizer};

        \draw[Arrow] (q_S_I_1-out) to[out=180, in=90] (kappa_I_1-in);
        \draw[Arrow] (s_tilde_I_sum_1) -- (q_S_I_1-in);
        \draw[Arrow] (kappa_tilde_I_1) to[out=90, in=-90] (s_tilde_I_sum_1);
        \draw[Arrow] (kappa_tilde_bar_I_1) to[out=90, in=0] (s_tilde_I_sum_1);
        \draw[Arrow] (kappa_I_1-out) -- (sum_I_1);
        \draw[Arrow] (x_I_1) to[out=90, in=-90] (kappa_tilde_I_1);
        \draw[Arrow] (x_Q_1) to[out=45, in=-90] (kappa_tilde_bar_I_1);
        \node at ($(s_tilde_I_sum_1) + (0.425,0.1875)$) {$-$};

            \draw pic[scale=1.5,rotate=180] (kappa_I_2) at ($(sum_I_2) + (0,1.6125)$) {DAC={$\kappa_\phi$}};
            \node[AnalogMultiplier] (kappa_tilde_I_2) at ($(x_I_2) + (0,1.75)$) {$\tilde{\kappa}_\phi$};
            \node[AnalogMultiplier] (kappa_tilde_bar_I_2) at ($(x_I_2) + (0.7,1.75)$) {$\bar{\tilde{\kappa}}_\phi$};
            \node[AnalogSum] (s_tilde_I_sum_2) at ($(kappa_tilde_I_2) + (0,1)$) {};
            \draw pic[rotate=0, yscale=1, xscale=-1] (q_S_I_2) at ($(s_tilde_I_sum_2) + (-1.375, 0)$) {quantizer};

            \draw[Arrow] (q_S_I_2-out) to[out=180, in=90] (kappa_I_2-in);
            \draw[Arrow] (s_tilde_I_sum_2) -- (q_S_I_2-in);
            \draw[Arrow] (kappa_tilde_I_2) to[out=90, in=-90] (s_tilde_I_sum_2);
            \draw[Arrow] (kappa_tilde_bar_I_2) to[out=90, in=0] (s_tilde_I_sum_2);
            \draw[Arrow] (kappa_I_2-out) -- (sum_I_2);
            \draw[Arrow] (x_I_2) to[out=90, in=-90] (kappa_tilde_I_2);
            \draw[Arrow] (x_Q_2) to[out=45, in=-90] (kappa_tilde_bar_I_2);
            \node at ($(s_tilde_I_sum_2) + (0.425,0.1875)$) {$-$};

            \draw pic[scale=1.5,rotate=180] (kappa_I_3) at ($(sum_I_3) + (0,1.6125)$) {DAC={$\kappa_\phi$}};
            \node[AnalogMultiplier] (kappa_tilde_I_3) at ($(x_I_3) + (0,1.75)$) {$\tilde{\kappa}_\phi$};
            \node[AnalogMultiplier] (kappa_tilde_bar_I_3) at ($(x_I_3) + (0.7,1.75)$) {$\bar{\tilde{\kappa}}_\phi$};
            \node[AnalogSum] (s_tilde_I_sum_3) at ($(kappa_tilde_I_3) + (0,1)$) {};
            \draw pic[rotate=0, yscale=1, xscale=-1] (q_S_I_3) at ($(s_tilde_I_sum_3) + (-1.375, 0)$) {quantizer};

            \draw[Arrow] (q_S_I_3-out) to[out=180, in=90] (kappa_I_3-in);
            \draw[Arrow] (s_tilde_I_sum_3) -- (q_S_I_3-in);
            \draw[Arrow] (kappa_tilde_I_3) to[out=90, in=-90] (s_tilde_I_sum_3);
            \draw[Arrow] (kappa_tilde_bar_I_3) to[out=90, in=0] (s_tilde_I_sum_3);
            \draw[Arrow] (kappa_I_3-out) --  (sum_I_3);
            \draw[Arrow] (x_I_3) to[out=90, in=-90] (kappa_tilde_I_3);
            \draw[Arrow] (x_Q_3) to[out=45, in=-90] (kappa_tilde_bar_I_3);
            \node at ($(s_tilde_I_sum_3) + (0.425,0.1875)$) {$-$};

            \draw pic[scale=1.5,rotate=180] (kappa_I_N) at ($(sum_I_N) + (0,1.6125)$) {DAC={$\kappa_\phi$}};
            \node[AnalogMultiplier] (kappa_tilde_I_N) at ($(x_I_N) + (0,1.75)$) {$\tilde{\kappa}_\phi$};
            \node[AnalogMultiplier] (kappa_tilde_bar_I_N) at ($(x_I_N) + (0.7,1.75)$) {$\bar{\tilde{\kappa}}_\phi$};
            \node[AnalogSum] (s_tilde_I_sum_N) at ($(kappa_tilde_I_N) + (0,1)$) {};
            \draw pic[rotate=0, yscale=1, xscale=-1] (q_S_I_N) at ($(s_tilde_I_sum_N) + (-1.375, 0)$) {quantizer};

            \draw[Arrow] (q_S_I_N-out) to[out=180, in=90] (kappa_I_N-in);
            \draw[Arrow] (s_tilde_I_sum_N) -- (q_S_I_N-in);
            \draw[Arrow] (kappa_tilde_I_N) to[out=90, in=-90] (s_tilde_I_sum_N);
            \draw[Arrow] (kappa_tilde_bar_I_N) to[out=115, in=-45] (s_tilde_I_sum_N);
            \draw[Arrow] (kappa_I_N-out) --  (sum_I_N);
            \draw[Arrow] (x_I_N) to[out=90, in=-90] (kappa_tilde_I_N);
            \draw[Arrow] (x_Q_N) to[out=45, in=-45] (kappa_tilde_bar_I_N);
            \node at ($(s_tilde_I_sum_N) + (0.425,0.1875)$) {$-$};

        \draw pic[scale=1.5,rotate=0] (kappa_Q_1) at ($(sum_Q_1) + (0,-1.6125)$) {DAC={$\kappa_\phi$}};
        \node[AnalogMultiplier] (kappa_tilde_Q_1) at ($(x_Q_1) + (0,-1.75)$) {$\tilde{\kappa}_\phi$};
        \node[AnalogMultiplier] (kappa_tilde_bar_Q_1) at ($(x_Q_1) + (0.7,-1.75)$) {$\bar{\tilde{\kappa}}_\phi$};
        \node[AnalogSum] (s_tilde_Q_sum_1) at ($(kappa_tilde_Q_1) + (0,-1)$) {};
        \draw pic[rotate=0, yscale=1, xscale=-1] (q_S_Q_1) at ($(s_tilde_Q_sum_1) + (-1.375, 0)$) {quantizer};

        \draw[Arrow] (q_S_Q_1-out) to[out=180, in=-90] (kappa_Q_1-in);
        \draw[Arrow] (s_tilde_Q_sum_1) -- (q_S_Q_1-in);
        \draw[Arrow] (kappa_tilde_Q_1) to[out=-90, in=90] (s_tilde_Q_sum_1);
        \draw[Arrow] (kappa_tilde_bar_Q_1) to[out=-90, in=0] (s_tilde_Q_sum_1);
        \draw[Arrow] (kappa_Q_1-out) -- (sum_Q_1);
        \draw[Arrow] (x_Q_1) to[out=-90, in=90] (kappa_tilde_Q_1);
        \draw[Arrow] (x_I_1) to[out=-45, in=90] (kappa_tilde_bar_Q_1);

            \draw pic[scale=1.5,rotate=0] (kappa_Q_2) at ($(sum_Q_2) + (0,-1.6125)$) {DAC={$\kappa_\phi$}};
            \node[AnalogMultiplier] (kappa_tilde_Q_2) at ($(x_Q_2) + (0,-1.75)$) {$\tilde{\kappa}_\phi$};
            \node[AnalogMultiplier] (kappa_tilde_bar_Q_2) at ($(x_Q_2) + (0.7,-1.75)$) {$\bar{\tilde{\kappa}}_\phi$};
            \node[AnalogSum] (s_tilde_Q_sum_2) at ($(kappa_tilde_Q_2) + (0,-1)$) {};
            \draw pic[rotate=0, yscale=1, xscale=-1] (q_S_Q_2) at ($(s_tilde_Q_sum_2) + (-1.375, 0)$) {quantizer};

            \draw[Arrow] (q_S_Q_2-out) to[out=180, in=-90] (kappa_Q_2-in);
            \draw[Arrow] (s_tilde_Q_sum_2) -- (q_S_Q_2-in);
            \draw[Arrow] (kappa_tilde_Q_2) to[out=-90, in=90] (s_tilde_Q_sum_2);
            \draw[Arrow] (kappa_tilde_bar_Q_2) to[out=-90, in=0] (s_tilde_Q_sum_2);
            \draw[Arrow] (kappa_Q_2-out) -- (sum_Q_2);
            \draw[Arrow] (x_Q_2) to[out=-90, in=90] (kappa_tilde_Q_2);
            \draw[Arrow] (x_I_2) to[out=-45, in=90] (kappa_tilde_bar_Q_2);

            \draw pic[scale=1.5,rotate=0] (kappa_Q_3) at ($(sum_Q_3) + (0,-1.6125)$) {DAC={$\kappa_\phi$}};
            \node[AnalogMultiplier] (kappa_tilde_Q_3) at ($(x_Q_3) + (0,-1.75)$) {$\tilde{\kappa}_\phi$};
            \node[AnalogMultiplier] (kappa_tilde_bar_Q_3) at ($(x_Q_3) + (0.7,-1.75)$) {$\bar{\tilde{\kappa}}_\phi$};
            \node[AnalogSum] (s_tilde_Q_sum_3) at ($(kappa_tilde_Q_3) + (0,-1)$) {};
            \draw pic[rotate=0, yscale=1, xscale=-1] (q_S_Q_3) at ($(s_tilde_Q_sum_3) + (-1.375, 0)$) {quantizer};

            \draw[Arrow] (q_S_Q_3-out) to[out=180, in=-90] (kappa_Q_3-in);
            \draw[Arrow] (s_tilde_Q_sum_3) -- (q_S_Q_3-in);
            \draw[Arrow] (kappa_tilde_Q_3) to[out=-90, in=90] (s_tilde_Q_sum_3);
            \draw[Arrow] (kappa_tilde_bar_Q_3) to[out=-90, in=0] (s_tilde_Q_sum_3);
            \draw[Arrow] (kappa_Q_3-out) -- (sum_Q_3);
            \draw[Arrow] (x_Q_3) to[out=-90, in=90] (kappa_tilde_Q_3);
            \draw[Arrow] (x_I_3) to[out=-45, in=90] (kappa_tilde_bar_Q_3);

            \draw pic[scale=1.5,rotate=0] (kappa_Q_N) at ($(sum_Q_N) + (0,-1.6125)$) {DAC={$\kappa_\phi$}};
            \node[AnalogMultiplier] (kappa_tilde_Q_N) at ($(x_Q_N) + (0,-1.75)$) {$\tilde{\kappa}_\phi$};
            \node[AnalogMultiplier] (kappa_tilde_bar_Q_N) at ($(x_Q_N) + (0.7,-1.75)$) {$\bar{\tilde{\kappa}}_\phi$};
            \node[AnalogSum] (s_tilde_Q_sum_N) at ($(kappa_tilde_Q_N) + (0,-1)$) {};
            \draw pic[rotate=0, yscale=1, xscale=-1] (q_S_Q_N) at ($(s_tilde_Q_sum_N) + (-1.375, 0)$) {quantizer};

            \draw[Arrow] (q_S_Q_N-out) to[out=180, in=-90] (kappa_Q_N-in);
            \draw[Arrow] (s_tilde_Q_sum_N) -- (q_S_Q_N-in);
            \draw[Arrow] (kappa_tilde_Q_N) to[out=-90, in=90] (s_tilde_Q_sum_N);
            \draw[Arrow] (kappa_tilde_bar_Q_N) to[out=-115, in=45] (s_tilde_Q_sum_N);
            \draw[Arrow] (kappa_Q_N-out) -- (sum_Q_N);
            \draw[Arrow] (x_Q_N) to[out=-90, in=90] (kappa_tilde_Q_N);
            \draw[Arrow] (x_I_N) to[out=-45, in=45] (kappa_tilde_bar_Q_N);

\end{tikzpicture}

%% file: figures/tf_lfs.tex
\begin{tikzpicture}
    \pgfplotsset{
        grid=both, 
    }
    \pgfplotsset{ylabel near ticks, xlabel near ticks}
    \pgfplotsset{ 
        legend cell align={left},
        legend style={
            at={(0.99,0.99)},
            anchor=north east,
            fill opacity=1,
            draw opacity=1,
            text opacity=1,
            nodes={scale=0.8, transform shape},
        }
    }

    \begin{axis}[
        name=plot2,
        xlabel={$f T_s$},
        ylabel={$\left|G_N(i 2\pi f)\right|$ dB},
        yminorgrids,
        ymajorgrids,
        xminorgrids,
        xmajorgrids,
        xmin=0,
        xmax=0.5,
        ymin=0,
        ymax=100,
        width=0.95\columnwidth,
        height=\columnwidth/2,
        xtick = {0, 0.0625, 0.125, 0.1875, 0.25, 0.3125, 0.375, 0.4375, 0.5},
        xticklabels = {$0$, ,$1/8$, ,$1/4$, ,$3/8$, , $1/2$},
        compat=1.16, 
        cycle list={
        {black,mark=},
        {blue,mark=},
        {olive,mark=},
        {red,mark=}
        },
        ]

        \addplot+[mark=, color=olive, thick] table[col sep=comma, x=f_lp_2, y=tf_lp_2] {./figures/tf_lf.csv};
        \addplot+[mark=, color=red, thick] table[col sep=comma, x=f_lp_1, y=tf_lp_1] {./figures/tf_lf.csv};
        \addplot+[mark=, color=black, thick] table[col sep=comma, x=f_lp_0, y=tf_lp_0] {./figures/tf_lf.csv};

        \addplot+[mark=, color=olive, thick, densely dashed] table[col sep=comma, x=f_2, y=tf_2] {./figures/tf_lf.csv};
        \addplot+[mark=, color=red, thick, densely dashed] table[col sep=comma, x=f_1, y=tf_1] {./figures/tf_lf.csv};
        \addplot+[mark=, color=black, thick, densely dashed] table[col sep=comma, x=f_0, y=tf_0] {./figures/tf_lf.csv};

        \addlegendentry{$N=2, f_n=0.45 f_s$}
        \addlegendentry{$N=4, f_n=0.3 f_s$}
        \addlegendentry{$N=8, f_n=0.15 f_s$}
                
    \end{axis}
\end{tikzpicture}

%% file: figures/psd.tex
\begin{tikzpicture}
    \pgfplotsset{
        grid=both, 
        minor grid style = {densely dotted}, 
        major grid style = {densely dotted}
    }
    \pgfplotsset{ylabel near ticks, xlabel near ticks}
    \pgfplotsset{ 
        legend cell align={left},
        legend style={
            at={(0.985,0.98)},
            anchor=north east,
            fill opacity=0.25,
            draw opacity=1,
            text opacity=1,
            nodes={scale=0.8, transform shape}
        }
    }

    \begin{axis}[
        name=plot2,
        yminorgrids,
        ymajorgrids,
        xminorgrids,
        xmajorgrids,
        xmin=0,
        xmax=0.5,
        ymin=-130,
        width=0.95\columnwidth,
        height=\columnwidth/2.3,
        xtick = {0, 0.0625, 0.125, 0.1875, 0.25, 0.3125, 0.375, 0.4375, 0.5},
        xticklabels = {$0$, ,$1/8$, ,$1/4$, ,$3/8$, , $1/2$},
        xticklabels = {},
        compat=1.16, 
       cycle list={
        {black,mark=},
        {blue,mark=},
        {olive,mark=},
        {red,mark=}
        },
        ]

        \addplot+[mark=, color=black] table[col sep=comma, x=f_fp_0, y expr={\thisrow{fp_0} + 69}] {./figures/psd_8_4.csv};
        \addplot+[mark=, color=red] table[col sep=comma, x=f_fp_1, y expr={\thisrow{fp_1} + 69}] {./figures/psd_8_4.csv};
        \addplot+[mark=, color=black] table[col sep=comma, x=f_fp_2, y expr={\thisrow{fp_2} + 69}] {./figures/psd_8_4.csv};
        \addplot+[mark=, color=red] table[col sep=comma, x=f_fp_3, y expr={\thisrow{fp_3} + 69}] {./figures/psd_8_4.csv};
        
        \addplot+[mark=, color=blue] table[col sep=comma, x=f_bb, y expr={\thisrow{bb} + 69}] {./figures/psd_8_4.csv};
        
    \end{axis}

    \begin{axis}[
        name=plot1,
        at=(plot2.below south), anchor=above north,
        yminorgrids,
        ymajorgrids,
        xminorgrids,
        xmajorgrids,
        xmin=0,
        xmax=0.5,
        ymin=-107,
        width=0.95\columnwidth,
        height=\columnwidth/2.3,
        xtick = {0, 0.0625, 0.125, 0.1875, 0.25, 0.3125, 0.375, 0.4375, 0.5},
        xticklabels = {$0$, ,$1/8$, ,$1/4$, ,$3/8$, , $1/2$},
        xticklabels = {},
        compat=1.16, 
       cycle list={
        {black,mark=},
        {blue,mark=},
        {olive,mark=},
        {red,mark=}
        },
        ylabel=dBFS,
        ]

        \addplot+[mark=, color=black] table[col sep=comma, x=f_fp_0, y expr={\thisrow{fp_0} + 67}] {./figures/psd_6_4.csv};
        \addplot+[mark=, color=olive] table[col sep=comma, x=f_fp_1, y expr={\thisrow{fp_1} + 67}] {./figures/psd_6_4.csv};
        \addplot+[mark=, color=black] table[col sep=comma, x=f_fp_2, y expr={\thisrow{fp_2} + 67}] {./figures/psd_6_4.csv};
        \addplot+[mark=, color=olive] table[col sep=comma, x=f_fp_3, y expr={\thisrow{fp_3} + 67}] {./figures/psd_6_4.csv};
        
        \addplot+[mark=, color=blue] table[col sep=comma, x=f_bb, y expr={\thisrow{bb} + 67}] {./figures/psd_6_4.csv};
        
    \end{axis}

    \begin{axis}[
        name=plot3,
        at=(plot1.below south), anchor=above north,
        xlabel={$fT_s$},
        yminorgrids,
        ymajorgrids,
        xminorgrids,
        xmajorgrids,
        xmin=0,
        xmax=0.5,
        ymin=-161,
        width=0.95\columnwidth,
        height=\columnwidth/2.3,
        xtick = {0, 0.0625, 0.125, 0.1875, 0.25, 0.3125, 0.375, 0.4375, 0.5},
        xticklabels = {$0$, ,$1/8$, ,$1/4$, ,$3/8$, , $1/2$},
        ]
        
        \addplot[mark=, color=black] table[col sep=comma, x=f_fp_0, y expr={\thisrow{fp_0} + 65}] {./figures/psd_6_8.csv};
        \addplot[mark=, color=orange] table[col sep=comma, x=f_fp_1, y expr={\thisrow{fp_1} + 65}] {./figures/psd_6_8.csv};
        \addplot[mark=, color=black] table[col sep=comma, x=f_fp_2, y expr={\thisrow{fp_2} + 65}] {./figures/psd_6_8.csv};
        \addplot[mark=, color=orange] table[col sep=comma, x=f_fp_3, y expr={\thisrow{fp_3} + 65}] {./figures/psd_6_8.csv};
        \addplot[mark=, color=black] table[col sep=comma, x=f_fp_4, y expr={\thisrow{fp_4} + 65}] {./figures/psd_6_8.csv};
        \addplot[mark=, color=orange] table[col sep=comma, x=f_fp_5, y expr={\thisrow{fp_5} + 65}] {./figures/psd_6_8.csv};
        \addplot[mark=, color=black] table[col sep=comma, x=f_fp_6, y expr={\thisrow{fp_6} + 65}] {./figures/psd_6_8.csv};
        \addplot[mark=, color=orange] table[col sep=comma, x=f_fp_7, y expr={\thisrow{fp_7} + 65}] {./figures/psd_6_8.csv};
        
        \addplot[mark=,color=blue] table[col sep=comma, x=f_bb, y expr={\thisrow{bb} + 65}] {./figures/psd_6_8.csv};

    \end{axis}
        
\end{tikzpicture}

%% file: figures/hist_FS.tex
\begin{tikzpicture}
    \pgfplotsset{
        grid=both, 
    }
    \pgfplotsset{ylabel near ticks, xlabel near ticks}
    \pgfplotsset{ 
        legend cell align={left},
        legend style={
            at={(0.985,0.98)},
            anchor=north east,
            fill opacity=0.25,
            draw opacity=1,
            text opacity=1,
            nodes={scale=0.8, transform shape}
        }
    }

\definecolor{darkgray176}{RGB}{176,176,176}
\definecolor{lightgray204}{RGB}{204,204,204}

\begin{axis}[
name=plot4,
at=(plot3.below south), anchor=above north,
width=0.95\columnwidth,
height=\columnwidth/2,
legend cell align={left},
legend style={
  fill opacity=0.8,
  draw opacity=1,
  text opacity=1,
  at={(0.01,0.97)},
  anchor=north west,
  draw=lightgray204
},
tick align=outside,
tick pos=left,
x grid style={darkgray176},
xlabel={SNR dB},
xmin=-4, xmax=2,
xtick style={color=black},
y grid style={darkgray176},
ymin=0, ymax=44,
ytick style={color=black}
]

\draw[draw=blue,fill=blue] (axis cs:-3.62619059305403,0) rectangle (axis cs:-3.36319891774028,5);
\addlegendimage{ybar,ybar legend,draw=blue,fill=blue}
\addlegendentry{Bandpass}
\draw[draw=blue,fill=blue] (axis cs:-3.36319891774028,0) rectangle (axis cs:-3.10020724242652,0);
\draw[draw=blue,fill=blue] (axis cs:-3.10020724242652,0) rectangle (axis cs:-2.83721556711277,8);
\draw[draw=blue,fill=blue] (axis cs:-2.83721556711277,0) rectangle (axis cs:-2.57422389179901,6);
\draw[draw=blue,fill=blue] (axis cs:-2.57422389179901,0) rectangle (axis cs:-2.31123221648526,16);
\draw[draw=blue,fill=blue] (axis cs:-2.31123221648526,0) rectangle (axis cs:-2.0482405411715,14);
\draw[draw=blue,fill=blue] (axis cs:-2.0482405411715,0) rectangle (axis cs:-1.78524886585774,20);
\draw[draw=blue,fill=blue] (axis cs:-1.78524886585774,0) rectangle (axis cs:-1.52225719054399,22);
\draw[draw=blue,fill=blue] (axis cs:-1.52225719054399,0) rectangle (axis cs:-1.25926551523023,33);
\draw[draw=blue,fill=blue] (axis cs:-1.25926551523023,0) rectangle (axis cs:-0.996273839916477,27);
\draw[draw=blue,fill=blue] (axis cs:-0.996273839916477,0) rectangle (axis cs:-0.733282164602722,29);
\draw[draw=blue,fill=blue] (axis cs:-0.733282164602722,0) rectangle (axis cs:-0.470290489288966,29);
\draw[draw=blue,fill=blue] (axis cs:-0.470290489288966,0) rectangle (axis cs:-0.20729881397521,15);
\draw[draw=blue,fill=blue] (axis cs:-0.20729881397521,0) rectangle (axis cs:0.0556928613385459,12);
\draw[draw=blue,fill=blue] (axis cs:0.0556928613385459,0) rectangle (axis cs:0.318684536652301,13);
\draw[draw=blue,fill=blue] (axis cs:0.318684536652301,0) rectangle (axis cs:0.581676211966057,2);
\draw[draw=blue,fill=blue] (axis cs:0.581676211966057,0) rectangle (axis cs:0.844667887279813,2);
\draw[draw=blue,fill=blue] (axis cs:0.844667887279813,0) rectangle (axis cs:1.10765956259357,0);
\draw[draw=blue,fill=blue] (axis cs:1.10765956259357,0) rectangle (axis cs:1.37065123790732,1);

\draw[draw=red, opacity=0.8,fill=red] (axis cs:-1.82533651911319,0) rectangle (axis cs:-1.64443147197269,2);
\addlegendimage{ybar,ybar legend,draw=red,fill=red}
\addlegendentry{Low-pass}
\draw[draw=red, opacity=0.8,fill=red] (axis cs:-1.64443147197269,0) rectangle (axis cs:-1.46352642483219,1);
\draw[draw=red, opacity=0.8,fill=red] (axis cs:-1.46352642483219,0) rectangle (axis cs:-1.28262137769168,4);
\draw[draw=red, opacity=0.8,fill=red] (axis cs:-1.28262137769168,0) rectangle (axis cs:-1.10171633055118,10);
\draw[draw=red, opacity=0.8,fill=red] (axis cs:-1.10171633055118,0) rectangle (axis cs:-0.920811283410679,12);
\draw[draw=red, opacity=0.8,fill=red] (axis cs:-0.920811283410679,0) rectangle (axis cs:-0.739906236270176,13);
\draw[draw=red, opacity=0.8,fill=red] (axis cs:-0.739906236270176,0) rectangle (axis cs:-0.559001189129674,22);
\draw[draw=red, opacity=0.8,fill=red] (axis cs:-0.559001189129674,0) rectangle (axis cs:-0.378096141989172,29);
\draw[draw=red, opacity=0.8,fill=red] (axis cs:-0.378096141989172,0) rectangle (axis cs:-0.197191094848669,30);
\draw[draw=red, opacity=0.8,fill=red] (axis cs:-0.197191094848669,0) rectangle (axis cs:-0.0162860477081672,23);
\draw[draw=red, opacity=0.8,fill=red] (axis cs:-0.0162860477081672,0) rectangle (axis cs:0.164618999432335,30);
\draw[draw=red, opacity=0.8,fill=red] (axis cs:0.164618999432335,0) rectangle (axis cs:0.345524046572837,18);
\draw[draw=red, opacity=0.8,fill=red] (axis cs:0.345524046572837,0) rectangle (axis cs:0.52642909371334,22);
\draw[draw=red, opacity=0.8,fill=red] (axis cs:0.52642909371334,0) rectangle (axis cs:0.707334140853842,11);
\draw[draw=red, opacity=0.8,fill=red] (axis cs:0.707334140853842,0) rectangle (axis cs:0.888239187994344,11);
\draw[draw=red, opacity=0.8,fill=red] (axis cs:0.888239187994344,0) rectangle (axis cs:1.06914423513485,10);
\draw[draw=red, opacity=0.8,fill=red] (axis cs:1.06914423513485,0) rectangle (axis cs:1.25004928227535,0);
\draw[draw=red, opacity=0.8,fill=red] (axis cs:1.25004928227535,0) rectangle (axis cs:1.43095432941585,4);
\draw[draw=red, opacity=0.8,fill=red] (axis cs:1.43095432941585,0) rectangle (axis cs:1.61185937655635,1);

\path [draw=black, very thick]
(axis cs:0,0)
--(axis cs:0,30) node[above]{Nominal};

\end{axis}

\node[rotate=90,xshift=-1.5cm, yshift=1cm] at (plot4.west) {Number of occurrences};

\begin{axis}[
name=plot5,
at=(plot4.below south), anchor=above north,
width=0.95\columnwidth,
height=\columnwidth/2,
legend cell align={left},
legend style={
  fill opacity=0.8,
  draw opacity=1,
  text opacity=1,
  at={(0.03,0.97)},
  anchor=north west,
  draw=lightgray204
},
tick align=outside,
tick pos=left,
x grid style={darkgray176},
xlabel={$\hat{f}_n / f_n$},
xmin=0.94, xmax=1.05,
xtick style={color=black},
y grid style={darkgray176},
ymin=0, ymax=35.7,
ytick style={color=black}
]
\draw[draw=blue,fill=blue] (axis cs:0.960433779411535,0) rectangle (axis cs:0.964790261834659,2);
\draw[draw=blue,fill=blue] (axis cs:0.964790261834659,0) rectangle (axis cs:0.969146744257783,3);
\draw[draw=blue,fill=blue] (axis cs:0.969146744257783,0) rectangle (axis cs:0.973503226680906,5);
\draw[draw=blue,fill=blue] (axis cs:0.973503226680906,0) rectangle (axis cs:0.97785970910403,6);
\draw[draw=blue,fill=blue] (axis cs:0.97785970910403,0) rectangle (axis cs:0.982216191527154,18);
\draw[draw=blue,fill=blue] (axis cs:0.982216191527154,0) rectangle (axis cs:0.986572673950278,21);
\draw[draw=blue,fill=blue] (axis cs:0.986572673950278,0) rectangle (axis cs:0.990929156373402,28);
\draw[draw=blue,fill=blue] (axis cs:0.990929156373402,0) rectangle (axis cs:0.995285638796525,22);
\draw[draw=blue,fill=blue] (axis cs:0.995285638796525,0) rectangle (axis cs:0.999642121219649,28);
\draw[draw=blue,fill=blue] (axis cs:0.999642121219649,0) rectangle (axis cs:1.00399860364277,28);
\draw[draw=blue,fill=blue] (axis cs:1.00399860364277,0) rectangle (axis cs:1.0083550860659,23);
\draw[draw=blue,fill=blue] (axis cs:1.0083550860659,0) rectangle (axis cs:1.01271156848902,16);
\draw[draw=blue,fill=blue] (axis cs:1.01271156848902,0) rectangle (axis cs:1.01706805091214,11);
\draw[draw=blue,fill=blue] (axis cs:1.01706805091214,0) rectangle (axis cs:1.02142453333527,19);
\draw[draw=blue,fill=blue] (axis cs:1.02142453333527,0) rectangle (axis cs:1.02578101575839,8);
\draw[draw=blue,fill=blue] (axis cs:1.02578101575839,0) rectangle (axis cs:1.03013749818152,5);
\draw[draw=blue,fill=blue] (axis cs:1.03013749818152,0) rectangle (axis cs:1.03449398060464,6);
\draw[draw=blue,fill=blue] (axis cs:1.03449398060464,0) rectangle (axis cs:1.03885046302776,1);
\draw[draw=blue,fill=blue] (axis cs:1.03885046302776,0) rectangle (axis cs:1.04320694545089,2);
\end{axis}

\end{tikzpicture}

%% file: figures/quadrature_implementation.tex
\begin{circuitikz}[european voltages]
\node[op amp, yscale=-1] (amp_I) at (0,0) {\scalebox{1}[-1]{$A(s)$}};
\node[rground] at (amp_I.+) {};
\node[inner sep=0] (vgnd_I) at ($(amp_I.-) + (-0.5,0.)$) {};
\draw (vgnd_I) to[short, *-] (amp_I.-);
\draw (vgnd_I) to[short, *-] ++(0,-1.125) to[C, l_=$C$] ++ (2.875,0) -| (amp_I.out);
\draw ($(amp_I.out)+(1.5,0)$) node[anchor=south] {$v_{x_\ell}$} to[short,o-] (amp_I.out);

\node[plain amp, yscale=1, rotate=180] (quantizer_I) at ($(amp_I) + (0, 1.5)$) {};
\draw ($(quantizer_I) + (0.5, -0.25)$) -- ++(-0.25,0) -- ++(0, 0.5) -- ++(-0.25, 0);
\draw ($(quantizer_I) + (0.1875, 0)$) -- ++(0.125,0);
\node (fs) at ($(quantizer_I) + (1.375, 0)$) {};
\draw ($(fs) + (0,-0.1875)$ 
    -- ++(0.125, 0) 
    -- ++(0,0.25) 
    -- ++(0.125,0) 
    -- ++(0, -0.25) 
    -- ++(0.125, 0)
    -- ++(0, 0.25)
    -- ++(0.125, 0)
    -- ++ (0, -0.25)
    -- ++ (0.125, 0)
    -- ++ (0, 0.25)
    -- ++ (0.125, 0)
    -- ++ (0, -0.25)
    -- ++ (0.125,0 )
;
\draw[{Stealth[length=0.875mm]}-{Stealth[length=0.875mm]}] ($(fs) + (0, 0.1375)$) -- node[above] {\tiny $1/f_s$} ++(0.25,0); 

\draw[Arrow] (fs) -- node[above] {\tiny CLK} ++(-0.5125, 0);
\node[left] at ($(quantizer_I.-) + (-0.3125,0)$) {$\tilde{\kappa}_\phi$};
\node[left] at ($(quantizer_I.+) + (-0.3125,0)$) {$-\bar{\tilde{\kappa}}_\phi$};

\draw (amp_I.out) to[short, *-] (quantizer_I.-);

\node[op amp] (amp_Q) at (0,-6.25) {$A(s)$};
\node[rground] at (amp_Q.+) {};
\node[inner sep=0] (vgnd_Q) at ($(amp_Q.-) + (-0.5,-0.0)$) {};
\draw (vgnd_Q) to[short, *-] (amp_Q.-);
\draw (vgnd_Q) to[short, *-] ++(0,1.125) to[C, l=$C$] ++ (2.875,0) -| (amp_Q.out);
\draw ($(amp_Q.out)+(1.5,0)$) node[anchor=south] {$v_{\bar{x}_\ell}$} to[short,o-] (amp_Q.out);

\node[plain amp, yscale=-1, rotate=180] (quantizer_Q) at ($(amp_Q) + (0, -1.5)$) {};
\draw ($(quantizer_Q) + (0.5, -0.25)$) -- ++(-0.25,0) -- ++(0, 0.5) -- ++(-0.25, 0);
\draw ($(quantizer_Q) + (0.1875, 0)$) -- ++(0.125,0);
\node (fs) at ($(quantizer_Q) + (1.375, 0)$) {};
\draw ($(fs) + (0,-0.1875)$ 
    -- ++(0.125, 0) 
    -- ++(0,0.25) 
    -- ++(0.125,0) 
    -- ++(0, -0.25) 
    -- ++(0.125, 0)
    -- ++(0, 0.25)
    -- ++(0.125, 0)
    -- ++ (0, -0.25)
    -- ++ (0.125, 0)
;
\draw[{Stealth[length=0.875mm]}-{Stealth[length=0.875mm]}] ($(fs) + (0, 0.1375)$) -- node[above] {\tiny $1/f_s$} ++(0.25,0); 

\draw[Arrow] (fs) -- node[above] {\tiny CLK} ++(-0.5125, 0);
\node[left] at ($(quantizer_Q.-) + (-0.3125,0)$) {$\tilde{\kappa}_\phi$};
\node[left] at ($(quantizer_Q.+) + (-0.3125,0)$) {$\bar{\tilde{\kappa}}_\phi$};

\draw (amp_Q.out) to[short, *-] (quantizer_Q.-);

\draw (quantizer_I.out) -- ++ (-0.5,0) coordinate (s_I) to[R, l_=$R_{\kappa_\phi}$, -] (vgnd_I);
\draw (quantizer_Q.out) -- ++ (-0.5,0) coordinate (s_Q) to[R, l=$R_{\kappa_\phi}$,-] (vgnd_Q);

\draw (amp_Q.out) -- ++(0.5,0.5) coordinate (out_Q) -- ++(0,2.375) -- ++(-0.75,0.5) -- ++(-3,0) to[R, l=$-R_{\omega_n}$] ++(0,1.75) -- (vgnd_I);
\draw (amp_I.out) -- ++(0.5,-0.5) coordinate (out_I) -- ++(0,-2.375) -- ++(-0.75,-0.5) -- ++(-3,0) to[R, l_=$R_{\omega_n}$] ++(0,-1.75) -- (vgnd_Q);

\draw (quantizer_I.+) -- ++(1.0, 0) |- ($(amp_Q.out) + (0.5, -0.5)$) -- (amp_Q.out);
\draw (quantizer_Q.+) -- ++(0.75,0) |- ($(amp_I.out) + (0.5, 0.5)$) -- (amp_I.out);

\draw ($(vgnd_I) + (-3, 0.)$) node[anchor=south] {$v_{x_{\ell-1}}$} to[R,o-,l_=$R_{\beta}$] ++(2,0) -- (vgnd_I);
\draw ($(vgnd_I) + (-3, 0.75)$) node[anchor=south] {$v_{x_{\ell+1}}$} to[R,o-,l=$R_{\alpha}$] ++(2,0) -- (vgnd_I);

\draw ($(vgnd_Q) + (-3, -0.0)$) node[anchor=south] {$v_{\bar{x}_{\ell-1}}$} to[R,o-,l=$R_{\beta}$] ++(2,0) -- (vgnd_Q);
\draw ($(vgnd_Q) + (-3, -0.75)$) node[anchor=south] {$v_{\bar{x}_{\ell+1}}$} to[R,o-,l_=$R_{\alpha}$] ++(2,0) -- (vgnd_Q);

\end{circuitikz}

%% file: figures/op-amp_snr_vs_GBWP.tex
\begin{tikzpicture}
    \pgfplotsset{
        grid=both, 
    }
    \pgfplotsset{ylabel near ticks, xlabel near ticks}
    \pgfplotsset{ 
        legend cell align={left},
        legend style={
            at={(0.99,0.01)},
            anchor=south east,
            fill opacity=1,
            draw opacity=1,
            text opacity=1,
            nodes={scale=0.7, transform shape},
            legend columns=2,
        }
    }
    \begin{semilogxaxis}[
        xlabel={$k_A \omega_A / \left(f_n + \frac{\omega_{\mathcal{B}}}{4 \pi}\right)$},
        ylabel={SNR dB},
        yminorgrids,
        ymajorgrids,
        xminorgrids,
        xmajorgrids,
        xmin=1,
        xmax=1e3,
        ymin=-20,
        width=0.95\columnwidth,
        height=\columnwidth/2,
        compat=1.18,
        cycle list={
            {teal,mark=x},
            {red,mark=*},
            {olive,mark=o},
            {blue,mark=star},  
            {magenta,mark=|},
            {orange,mark=},
            {cyan,mark=},
            {purple,mark=+},
        },
        ]

        \addplot+[thick] table[col sep=comma,x=f,y=SNR] {./figures/snr_op-amp_2.00e+01_6_4.csv};
        \addlegendentry{$k_{\mathrm{DC}} = 20 \frac{\mathrm{OSR}}{\pi}$}

        \addplot+[thick] table[col sep=comma,x=f,y=SNR] {./figures/snr_op-amp_3.00e+01_6_4.csv};
        \addlegendentry{$k_{\mathrm{DC}} = 30 \frac{\mathrm{OSR}}{\pi}$}

        \addplot+[thick] table[col sep=comma,x=f,y=SNR] {./figures/snr_op-amp_5.00e+01_6_4.csv};
        \addlegendentry{$k_{\mathrm{DC}} = 50 \frac{\mathrm{OSR}}{\pi}$}

        \addplot+[thick] table[col sep=comma,x=f,y=SNR] {./figures/snr_op-amp_5.00e+02_6_4.csv};
        \addlegendentry{$k_{\mathrm{DC}} = 500 \frac{\mathrm{OSR}}{\pi}$}

        \addplot+[thick] table[col sep=comma,x=f,y=SNR] {./figures/snr_op-amp_1.00e+04_6_4.csv};
        \addlegendentry{$k_{\mathrm{DC}} = 10^4 \frac{\mathrm{OSR}}{\pi}$}

    \end{semilogxaxis}
\end{tikzpicture}

%% file: figures/op-amp_psd.tex
\begin{tikzpicture}
    \pgfplotsset{
        grid=both, 
        minor grid style = {densely dotted}, 
        major grid style = {densely dotted}
    }
    \pgfplotsset{ylabel near ticks, xlabel near ticks}
    \pgfplotsset{ 
        legend cell align={left},
        legend style={
            at={(0.01,0.99)},
            anchor=north west,
            fill opacity=1,
            draw opacity=1,
            text opacity=1,
            nodes={scale=0.7, transform shape}
        }
    }
    \begin{axis}[
        xlabel={MHz},
        ylabel=$\text{V} / \sqrt{\text{Hz}}$ dB,
        yminorgrids,
        ymajorgrids,
        xminorgrids,
        xmajorgrids,
        xmin=0,
        xmax=0.5,
        ymin=-180,
        width=0.95\columnwidth,
        height=\columnwidth/1.618,
        xtick = {0, 0.0625, 0.125, 0.1875, 0.25, 0.3125, 0.375, 0.4375, 0.5},
        xticklabels = {$0$, $134$, $268$, $403$, $537$, $671$, $805$, $940$, $1074$},
        ]
        \addplot[mark=, color=black] table[col sep=comma, x=f_fp_2, y=fp_2] {./figures/psd_6_4.csv};
        \addlegendentry{Wiener filter from \Fig{fig:psd}};
        
        \addplot[color=teal] table[col sep=comma,x=f_20_750,y expr={\thisrow{PSD_20_750} - 9}] {./figures/psd_op_amp_6_4.csv};
        \addlegendentry{$\kappa_A \omega_A = 750  \left(f_n + \frac{\omega_{\mathcal{B}}}{4 \pi}\right)$};
        
        \addplot[color=olive] table[col sep=comma,x=f_50_18,y expr={\thisrow{PSD_50_18} - 9}] {./figures/psd_op_amp_6_4.csv};
        \addlegendentry{$\kappa_A \omega_A = 18 \left(f_n + \frac{\omega_{\mathcal{B}}}{4 \pi}\right)$};

        \addplot[mark=, color=blue] table[col sep=comma,x=f_500_750,y expr={\thisrow{PSD_500_750} - 9}] {./figures/psd_op_amp_6_4.csv};
        \addlegendentry{$\kappa_A \omega_A = 750 \left(f_n + \frac{\omega_{\mathcal{B}}}{4 \pi}\right)$};

    \end{axis}
\end{tikzpicture}

%% file: figures/calibrated_filter.tex
\begin{tikzpicture}
    \pgfplotsset{
        grid=both, 
        minor grid style = {densely dotted}, 
        major grid style = {densely dotted}
    }
    \pgfplotsset{ylabel near ticks, xlabel near ticks}
    \pgfplotsset{ 
        legend cell align={left},
        legend style={
            at={(0.02,0.98)},
            anchor=north west,
            fill opacity=1.0,
            draw opacity=1,
            text opacity=1,
            nodes={scale=0.75, transform shape}
        }
    }

    \begin{axis}[
        name=plot1,
        at=(plot2.below south), anchor=above north,
        yminorgrids,
        ymajorgrids,
        xminorgrids,
        xmajorgrids,
        xmin=0,
        xmax=0.5,
        ymin=-90,
        width=0.95\columnwidth,
        height=\columnwidth/1.618,
        compat=1.16,
        cycle list={
            {black,mark=},
            {red,mark=},
            {olive,mark=},
            {blue,mark=},            
            {orange,mark=},
            {magenta,mark=},
            {cyan,mark=},
        },
        ylabel=$|\cdot|$ dB,
        xlabel=$\Omega / (2 \pi)$,
        ]
        \addplot+[mark=,thick] table[col sep=comma, x=f, y=h_0] {./figures/filter_plot_bode_op_amp_14_512.csv};
        \addlegendentry{$H_0(e^{i \Omega})$};
        \addplot+[mark=] table[col sep=comma, x=f, y=h_2] {./figures/filter_plot_bode_op_amp_14_512.csv};
        \addlegendentry{$H_1(e^{i \Omega})$};
        \addplot+[mark=] table[col sep=comma, x=f, y=h_3] {./figures/filter_plot_bode_op_amp_14_512.csv};
        \addlegendentry{$H_2(e^{i \Omega})$};
        \addplot+[mark=] table[col sep=comma, x=f, y=h_4] {./figures/filter_plot_bode_op_amp_14_512.csv};
        \addlegendentry{$H_3(e^{i \Omega})$};
        \addplot+[mark=] table[col sep=comma, x=f, y=h_5] {./figures/filter_plot_bode_op_amp_14_512.csv};
        \addlegendentry{$H_4(e^{i \Omega})$};
        \addplot+[mark=] table[col sep=comma, x=f, y=h_6] {./figures/filter_plot_bode_op_amp_14_512.csv};
        \addlegendentry{$H_5(e^{i \Omega})$};
        \addplot+[mark=] table[col sep=comma, x=f, y=h_7] {./figures/filter_plot_bode_op_amp_14_512.csv};
        \addlegendentry{$H_6(e^{i \Omega})$};
        
    \end{axis}

\end{tikzpicture}

%% file: figures/oscillator_AF.tex
\begin{tikzpicture}[node distance=1.25cm]
    \node[] (r_input) {$u_\ell(t)$};
    \node[AnalogMultiplier] (r_b) at ($(r_input)+(1.25,0)$) {$\beta$};
    \node[AnalogSum] (r_sum) at ($(r_b)+(1.125,0)$) {};
    \node[Analog] (r_int) at ($(r_sum) + (1,0)$) {$\int$};
    \node[AnalogBranch] (r_branch) at ($(r_int) + (1.375,0)$) {};

    \draw pic[rotate=180, yscale=-1] (r_quantizer) at ($(r_int) + (0, 1.75)$) {quantizer};
    \draw pic[scale=1.2,rotate=180] (r_kappa) at ($(r_sum) + (0,1)$) {DAC={$\kappa$}};
    \draw pic[scale=1.2,rotate=180] (r_kappa_bar) at ($(r_sum) + (-0.75,1)$) {DAC={$\bar{\kappa}$}};
    \node[AnalogBranch] (r_kappa_branch) at ($(r_kappa-in) + (0, 0.45)$) {};
    \node[AnalogMultiplier] (r_tilde) at ($(r_branch) + (0,1)$) {$\tilde{\kappa}$}; 
    \node[AnalogMultiplier] (r_tilde_bar) at ($(r_branch) + (0.6125,1)$) {$\bar{\tilde{\kappa}}$}; 
    \node[AnalogSum] (r_tilde_sum) at ($(r_quantizer-in) + (1.0,0)$) {};
    \node[above] at (r_kappa_branch) {$s_\ell[k]$};

    \draw[Arrow] (r_input) -- (r_b);
    \draw[Arrow] (r_b) -- (r_sum);
    \draw[Arrow] (r_sum) -- (r_int);
    \draw[] (r_int) -- node[above] {$x_\ell(t)$} (r_branch);
    
    \draw[Arrow] (r_branch) -- (r_tilde);
    \draw[Arrow] (r_tilde) -- (r_tilde_sum);
    \draw[Arrow] (r_tilde_bar) to[out=90, in=0] (r_tilde_sum);
    \draw[Arrow] (r_tilde_sum) -- node[below] {$\tilde{s}_\ell(t)$} (r_quantizer-in);
    \draw[] (r_quantizer-out) -- (r_kappa_branch);
    \draw[] (r_kappa_branch) -- (r_kappa-in);
    \draw[] (r_kappa_branch) to[out=180,in=90] (r_kappa_bar-in);
    \draw[Arrow] (r_kappa-out) -- (r_sum);

    \node[] (i_input) at ($(r_input) + (0,-2.125)$) {$\bar{u}_\ell(t)$};
    \node[AnalogMultiplier] (i_b) at ($(i_input)+(1.25,0)$) {$\beta$};
    \node[AnalogSum] (i_sum) at ($(i_b)+(1.125,0)$) {};
    \node[Analog] (i_int) at ($(i_sum) + (1,0)$) {$\int$};
    \node[AnalogBranch] (i_branch) at ($(i_int) + (1.375,0)$) {};

    \draw pic[rotate=180, yscale=-1] (i_quantizer) at ($(i_int) + (0, -1.75)$) {quantizer};
    \draw pic[scale=1.2,rotate=0] (i_kappa) at ($(i_sum) + (0,-1)$) {DAC={$\kappa$}};
    \draw pic[scale=1.2,rotate=0] (i_kappa_bar) at ($(i_sum) + (-0.75,-1)$) {DAC={$\bar{\kappa}$}};
    \node[AnalogBranch] (i_kappa_branch) at ($(i_kappa-in) + (0, -0.45)$) {};
    \node[AnalogMultiplier] (i_tilde) at ($(i_branch) + (0,-1)$) {$\tilde{\kappa}$}; 
    \node[AnalogMultiplier] (i_tilde_bar) at ($(i_branch) + (0.6125,-1)$) {$\bar{\tilde{\kappa}}$}; 
    \node[AnalogSum] (i_tilde_sum) at ($(i_quantizer-in) + (1.0,0)$) {};
    \node[below] at (i_kappa_branch) {$\bar{s}_\ell[k]$};

    \draw[Arrow] (i_branch) -- (i_tilde);
    \draw[Arrow] (i_tilde) -- (i_tilde_sum);
    \draw[Arrow] (i_tilde_bar) to[out=-90, in=0] (i_tilde_sum);
    \draw[Arrow] (i_tilde_sum) -- node[below] {$\bar{\tilde{s}}_\ell(t)$} (i_quantizer-in);
    \draw[] (i_quantizer-out) -- (i_kappa_branch);
    \draw[] (i_kappa_branch) -- (i_kappa-in);
    \draw[] (i_kappa_branch) to[out=180,in=-90] (i_kappa_bar-in);
    \draw[Arrow] (i_kappa-out) -- (i_sum);

    \node[AnalogMultiplier] (r_wp) at ($(r_int) + (0,-0.75)$) {$\omega_n$};
    \node[AnalogMultiplier] (i_wp) at ($(i_int) + (0,0.75)$) {$\omega_n$};

    \draw[Arrow] (i_input) -- (i_b);
    \draw[Arrow] (i_b) -- (i_sum);
    \draw[Arrow] (i_sum) -- (i_int);
    \draw[] (i_int) -- node[below] {$\bar{x}_\ell(t)$} (i_branch);

    \draw[Arrow] (r_wp) to[out=180, in=-90] (r_sum);
    \draw[Arrow] (r_branch) to[out=-90, in=0] (i_wp);
    \draw[Arrow] (i_branch) to[out=45,in=-90] (r_tilde_bar);
    \draw[Arrow] (r_kappa_bar-out) to[out=-90,in=135] (i_sum);

    \draw[Arrow] (i_wp) to[out=180, in=90] (i_sum);
    \draw[Arrow] (i_branch) to[out=90, in=0] (r_wp);
    \draw[Arrow] (r_branch) to[out=-45,in=90] (i_tilde_bar);
    \draw[Arrow] (i_kappa_bar-out) to[out=90,in=-135] (r_sum);

    \node[] at ($(r_sum) + (-0.1,-0.5)$) {$-$};
    \node[] at ($(r_sum) + (-0.5875,-0.325)$) {$-$};
    \node[] at ($(r_tilde_sum) + (0.5,0.125)$) {$-$};
\end{tikzpicture}